\documentclass{article}

\usepackage{arxiv}

\usepackage{bigfoot}
\usepackage{etoolbox}

\usepackage[utf8]{inputenc} 
\usepackage[T1]{fontenc}    
\usepackage[hidelinks]{hyperref}       
\usepackage{url}            
\usepackage{booktabs}       
\usepackage{amsfonts}       
\usepackage{amsmath}
\usepackage{amssymb}
\usepackage{nicefrac}       
\usepackage{microtype}      
\usepackage{cleveref}       
\usepackage{graphicx}
\usepackage[numbers,sort&compress,square]{natbib}
\usepackage{doi}

\usepackage{float}
\usepackage{graphicx}
\usepackage{siunitx}
\usepackage{subcaption}
\usepackage{pgfplots}
\usepackage{caption}
\usepackage{flafter}
\usepackage{comment}
\usepackage[acronym,nonumberlist]{glossaries}
\usepackage{multirow}
\usepackage{xcolor}
\pgfplotsset{compat=newest}

\makenoidxglossaries
\newglossaryentry{FOM}
{
    name=FOM,
    description={full-order model}
}
\newglossaryentry{ROM}
{
    name=ROM,
    description={reduced-order model}
}
\newglossaryentry{MOR}
{
    name=MOR,
    description={model order reduction}
}
\newglossaryentry{INS}
{
    name=INS,
    description={incompressible Navier-Stokes}
}
\newglossaryentry{AD}
{
    name=AD,
    description={advection-diffusion}
}
\newglossaryentry{CBRN}
{
    name=CBRN,
    description={chemical, biological, radiological, nuclear}
}
\newglossaryentry{POD}
{
    name=POD,
    description={proper orthogonal decomposition}
}
\newglossaryentry{SVD}
{
    name=SVD,
    description={singular value decomposition}
}
\newglossaryentry{PODG}
{
    name=PODG,
    description={proper orthogonal decomposition with subsequent Galerkin projection}
}
\newglossaryentry{PODI}
{
    name=PODI,
    description={proper orthogonal decomposition with subsequent interpolation}
}
\newglossaryentry{RBF}
{
    name=RBF,
    description={radial basis function}
}
\newglossaryentry{RB}
{
    name=RB,
    description={reduced basis}
}
\newglossaryentry{DEIM}
{
    name=DEIM,
    description={discrete empirical interpolation method}
}
\newglossaryentry{CFD}
{
    name=CFD,
    description={computational fluid dynamics}
}
\newglossaryentry{SUPG}
{
    name=SUPG,
    description={streamline upwind Petrov-Galerkin}
}

\newglossaryentry{PDE}
{
    name=PDE,
    description={partial differential equation}
}

\newglossaryentry{RANS}
{
    name=RANS,
    description={Reynolds-averaged Navier-Stokes}
}

\newglossaryentry{LES}
{
    name=LES,
    description={large eddy simulation}
}

\AtBeginDocument{
  \hypersetup{hidelinks}
}

\DeclareMathOperator{\diag}{diag}
\DeclareMathOperator{\spn}{span}

\newcommand{\mbQ}{\mathbf{Q}}
\newcommand{\mbu}{\mathbf{u}}

\newcommand{\mbuh}{\mathbf{u}^\mathrm{h}}

\newcommand{\mbn}{\mathbf{n}}
\newcommand{\mbgamma}{\boldsymbol{\gamma}}
\newcommand{\mbrho}{\boldsymbol{\rho}}
\newcommand{\rank}{r}

\newcommand{\p}{p}

\newcommand{\mbph}{\mathbf{p}^\mathrm{h}}

\newcommand{\mbg}{\mathbf{g}}
\newcommand{\mbf}{\mathbf{f}}
\newcommand{\mbc}{\mathbf{c}}
\newcommand{\mbnull}{\mathbf{0}}

\newcommand{\wdir}{w_{\mathrm{d}}}
\newcommand{\wint}{w_{\mathrm{i}}}

\newcommand{\mbur}{\hat{\mathbf{u}}}
\newcommand{\mbpr}{\hat{\mathbf{p}}}
\newcommand{\redA}{\hat{\mathbf{A}}}
\newcommand{\redC}{\hat{\mathbf{C}}}
\newcommand{\redB}{\hat{\mathbf{B}}}
\newcommand{\redf}{\hat{\mathbf{f}}}
\newcommand{\redg}{\hat{\mathbf{g}}}

\newcommand{\mbN}{\mathbf{N}}
\newcommand{\mbL}{\mathbf{L}}

\newcommand{\snap}{\mathbf{S}}
\newcommand{\s}{\mathbf{s}}
\newcommand{\subspace}{\mathcal{V}}
\newcommand{\projmatrix}{\mathbf{V}}
\newcommand{\projmatrixp}{\mathbf{P}}

\newcommand{\romdim}{N_{\mathrm{rb}}}
\newcommand{\romdimp}{N_{\mathrm{rb,p}}}
\newcommand{\fomdim}{N_\mathrm{h}}
\newcommand{\fomdimp}{N_{\mathrm{h,p}}}
\newcommand{\deimdim}{N_{\mathrm{DEIM}}}
\newcommand{\params}{\boldsymbol{\mu}}
\newcommand{\paramstrain}{\boldsymbol{\mu}_\mathrm{s}}
\newcommand{\trainset}{\Xi_\mathrm{s}}
\newcommand{\Rey}{\mathrm{Re}}

\newcommand{\inflowBoundary}{\Gamma_{-}}
\newcommand{\characteristicBoundary}{\Gamma_{0}}
\newcommand{\outflowBoundary}{\Gamma_{+}}

\newcommand{\normal}{\textbf{n}}

\newcommand{\nsnap}{N_{\mathrm{s}}}
\newcommand{\R}{\mathbb{R}}

\newcommand{\mbx}{\mathbf{x}}
\newcommand{\parameterInitial}{m_\mathrm{I}(\mbx)}
\newcommand{\parameterSource}{m_\mathrm{C}(\mbx)}

\DeclareNewFootnote{AAffil}[arabic]
\DeclareNewFootnote{ANote}[fnsymbol]

\makeatletter
\patchcmd\maketitle{\def\@makefnmark{\rlap{\@textsuperscript{\normalfont\@thefnmark}}}}{}{}{}
\makeatother

\makeatletter
\def\thanksAAffil#1{
  \footnotemarkAAffil\protected@xdef\@thanks{\@thanks%
        \protect\footnotetextAAffil[\the \c@footnoteAAffil]{#1}}%
}
\def\thanksANote#1{%
  \footnotemarkANote%
  \protected@xdef\@thanks{\@thanks%
        \protect\footnotetextANote[\the \c@footnoteANote]{#1}}%
}
\makeatother

\title{Intrusive and Non-Intrusive Model Order Reduction for Airborne Contaminant Transport: Comparative Analysis and Uncertainty Quantification}

\usepackage{authblk}

\newbox{\orcid}\sbox{\orcid}{\includegraphics[scale=0.06]{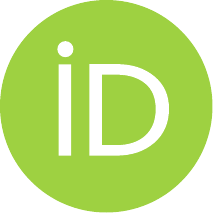}} 
\author[1]{%
	\href{https://orcid.org/0009-0006-8215-6093}{\usebox{\orcid}\hspace{1mm}Lisa K\"uhn\thanks{\texttt{lisa.kuehn@dlr.de}}}%
	}
\author[1]{%
	\href{https://orcid.org/0000-0001-8435-6466}{\usebox{\orcid}\hspace{1mm}Jacopo Bonari}%
	}
\author[1]{%
	\href{https://orcid.org/0000-0002-2814-0027}{\usebox{\orcid}\hspace{1mm}Max {v. Danwitz}}%
}
\author[1,2]{%
	\href{https://orcid.org/0000-0002-8820-466X}{\usebox{\orcid}\hspace{1mm}Alexander Popp}%
}
\affil[1]{German Aerospace Center (DLR), Institute for the Protection of Terrestrial Infrastructures, Rathausallee 12, 53757 St. Augustin, Germany}
\affil[2]{University of the Bundeswehr Munich, Institute for Mathematics and Computer-Based Simulation, Werner-Heisenberg-Weg 39, 85577 Neubiberg, Germany}

\renewcommand{\shorttitle}{MOR for Airborne Contaminant Transport} 

\hypersetup{
pdftitle={intrusive-and-nonintrusive-mor-for-airborne-contaminant-transport-comparative-analysis-and-uq},
pdfsubject={cs.CE},
pdfauthor={Lisa K\"uhn, Jacopo Bonari, Max v. Danwitz, Alexander Popp}, 
pdfkeywords={reduced-order modeling, airborne contaminant transport, urban physics, intrusive and non-intrusive MOR, proper orthogonal decomposition.}
}

\begin{document}
\maketitle

\begin{abstract}
Numerical simulations of contaminant dispersion after a gas leakage incident at a chemical plant can provide valuable insights for both emergency response and preparedness. High-fidelity simulation approaches combine incompressible Navier-Stokes (INS) equations with advection-diffusion processes to model wind and concentration field. However, their computational cost increases rapidly for complex geometries and extended domains like urban environments. This renders them unfeasible in time-critical or multi-query "what-if" scenarios, making model order reduction (MOR) techniques particularly valuable for enabling fast, accurate predictions. This study describes a MOR-based, application-driven workflow for uncertainty-aware contaminant concentration prediction, demonstrated for a two-dimensional benchmark system. Starting with the model selection, relevant comparison criteria and trade-offs are discussed for well-established proper orthogonal decomposition based (non-)intrusive methods, applied to the computationally more demanding parametric INS problem. They include accuracy, computational efficiency, data requirements, and extrapolation capability. Based on these insights, a non-intrusive parametric reduced-order model is constructed that enables accelerated Monte Carlo simulations for estimating the influence of potential wind measurement uncertainties on the spatio-temporal concentration field. This enables extracting valuable information for local consequence analyses and evacuation planning. Simulation results are furthermore interactively visualized in a dashboard and can serve as a building block within broader decision-support systems.
\end{abstract}

\keywords{Reduced-order modeling \and Airborne contaminant transport \and Urban physics \and Intrusive and non-intrusive MOR \and Proper orthogonal decomposition}

\section{Introduction}\label{sec:intro}

CBRN events, i.e., chemical, biological, radiological and nuclear events, caused either by accident, sabotage, or terrorism pose high risks to persons and critical infrastructures. Among CBRN threats, a major role is played by chemical incidents and the subsequent atmospheric release of dangerous contaminants. In this context, the U.S. Chemical Safety Board alone reports more detailed investigations for 80 chemical incidents in the U.S. within the last five years~\cite{Owens.2025}, with hundreds of other incidents reported throughout the last decades around the world as analyzed in~\cite{Bene.2024,Gomez.2008,comingcleaninc.2023} and references therein. Computational modeling and simulation can help investigating the spatio-temporal contaminant dispersion in such cases. For this purpose, tools such as~\emph{ADMS}~\cite{Carruthers.1994},~\emph{ALOHA}~\cite{jones.2013}, and~\emph{AERMOD}~\cite{Cimorelli.2005} have been introduced to estimate the contaminant concentration field based on a Gaussian plume model~\cite{holzbecher.2011,Snoun.2023}. They assume a continuous point source for the contaminant and calculate the downwind concentration under different simplification assumptions, first and foremost that the wind field is uniform and constant, and the variations in topography are negligible~\cite{Stockie.2011}.
Although already successfully applied in different analysis scenarios as, e.g.,~in~\cite{Schneider.2025,GarciaDiaz.2012,Law.2019,Liu.2024,Sheng.2024,Shi.2023} for industrial settings, these estimates -- in their basic form -- do not include the influence of obstacles on the concentration field as it is expected in case the contaminant spreads across built-up areas. To overcome this and other limitations of plume models, more complex simulation settings can be employed~\cite{Patnaik.2012,Hanna.2009,Shen.2020,Yuan.2022}. They can be broadly categorized into the three main families of Lagrangian, Eulerian and mixed models. An exhaustive, though unfortunately not up-to-date list of models employed in Europe can be found in~\cite{Kukkonen.2012}. Within the field of Eulerian computational fluid dynamics (\gls{CFD}) models, the airborne contaminant transport can be modeled by employing steady-state incompressible Navier-Stokes (\gls{INS}) equations to estimate the wind field and then solve a related advection-diffusion (\gls{AD}) problem, employing the computed wind field solution as advection field~\cite{Khamlich.2023,Danwitz.2024,Hinsen.2024}. Such models can offer a high accuracy and could theoretically be applied for predicting the contaminant transport in case of an incident or in preparatory "what-if" case scenario studies. However, their increased computational expense often prohibits the use for time-critical applications like chemical incidents and multi-query analysis, where the delivery of near real-time results is fundamental. To mitigate this problem, a promising solution is offered by the application of model order reduction (\gls{MOR}) techniques, aiming for fast yet accurate predictions. For this purpose, a reduced-order model (\gls{ROM}) is derived that provides a good approximation of the full-order model (\gls{FOM}), for example a high-fidelity \gls{CFD}-based model. This \gls{ROM} requires only a fraction of the computational resources compared to the \gls{FOM} and can thus be employed instead of the \gls{FOM}. For the considered application, the usage of \gls{MOR} techniques is particularly well-suited, since the approximation of parametric partial differential equations (\gls{PDE}s) is an active research area and a major field of application of \gls{MOR} methods~\cite{Rozza.2022, Quarteroni.2015, BennerVol1.2021, Benner.2020}.

In the following paragraphs, a literature review on the subject is offered, that starts with a general introduction to MOR for parametric~\gls{PDE}s before moving on to reduced models and uncertainty quantification for the specific application of airborne contaminant transport.

To enable fast evaluation times of the \gls{ROM}, a strict online-offline decomposition has to be employed. This includes an offline phase where snapshot solutions of the \gls{FOM} are taken for several parameter value samples. Based on these, the \gls{ROM} is set up such that evaluations during the online phase are completely independent of the \gls{FOM} and can hence be performed in very short time. Such reduction methods can be distinguished into two types: intrusive and non-intrusive methods. Intrusive methods require access to \gls{FOM} operators and intrude the numerical model code, e.g., a finite element formulation. An overview of common intrusive methods can be found in~\cite{Benner.2020}. Well-established methods for parametric \gls{PDE}s perform a projection of \gls{FOM} operators onto a reduced basis (\gls{RB}) that is set up based on \gls{FOM} solution snapshots. For setting up the reduced basis either a greedy algorithm or proper orthogonal decomposition (\gls{POD}) can be used~\cite{Quarteroni.2015,Pichi.2022,Quarteroni.2014,Hesthaven.2016}.
Non-intrusive methods, on the other hand, only rely on the high-fidelity data -- either from \gls{FOM} simulations or measurements -- and thus require no access to \gls{FOM} code. As for the intrusive methods, a common approach involves identifying a reduced, also called latent, space representation based on which predictions for the full-field solution can be derived. Such a reduced representation is employed by the more classical methods of \gls{POD} with interpolation (\gls{PODI})~\cite{BuiThanh.2003,Tezzele.2022,Yu.2019,Czech.2022,Regazzoni.2024,Xiang.2021}, dynamic mode decomposition~\cite{Kutz.2016,Tezzele.2022}, as well as neural-network based reduction methods like various types of autoencoders~\cite{Maulik.2021,Xiang.2021,Seo.2022,Fu.2023}. A broader overview on non-intrusive machine-learning based \gls{ROM}s, as well as more recent developments in the field, can be found in~\cite{Quarteroni.2025,Chinesta.2025}. \\
Focusing on the very broadly applied intrusive POD with Galerkin projection (\gls{PODG}) and non-intrusive \gls{PODI}, several advantages and disadvantages are identified in literature.
Intrusive \gls{PODG} is based on physical relationships and therefore often provides reliable and physically meaningful results~\cite{Conti.2024}. This can be even the case for parameter values slightly outside the trained parameter range, if the corresponding solution is sufficiently similar to those observed during training~\cite{Czech.2022,Girfoglio.2022}.
However, \gls{PODG} often comes with a lower speed-up~\cite{Czech.2022,Padula.2024} and a limited flexibility in terms of full-order simulators which have to be accessible for implementing the projection~\cite{Conti.2024}. Moreover, hyper-reduction methods may be required to maintain computational efficiency in the presence of nonlinearities or nonaffine parameter dependencies~\cite{Ballarin.2015,Cicci.2022, Quarteroni.2015}.
Further methods have been developed to circumvent stability issues of \gls{PODG} which may particularly arise for convection-dominated problems~\cite{Xiang.2021,Chen.2021,Ballarin.2015,Huang.2018}. These methods, for instance, rely on Petrov-Galerkin projection~\cite{Carlberg.2017,Grimberg.2020,Xiao.2013}, $L^1$-minimization~\cite{Abgrall.2018}, and supremizer stabilization~\cite{Ballarin.2015}.
Non-intrusive \gls{PODI}, on the other hand, can also be applied in case of a closed-source numerical solver for the \gls{FOM} as it is the case for commercial solvers. On the downside, it is known to need a lot of training data for approximating the \gls{FOM} behavior, leading to a heavier offline phase~\cite{Yu.2019,Czech.2022,Vinuesa.2022}. Although the reduced basis is based on \gls{FOM}-simulations, the approximation of the POD coefficients -- including either interpolation or regression -- lacks physical knowledge if not explicitly included as in~\cite{Chen.2021,Dave.2025}. Furthermore, applied regression methods may be vulnerable to overfitting or sensitive hyper-parameters~\cite{Czech.2022}. 
One drawback of both \gls{PODI} and \gls{PODG} is the linear subspace that is spanned by \gls{POD}-based approaches, making it difficult to adequately represent problems featuring a slowly decaying Kolmogorov n-width \cite{Regazzoni.2024}, i.e., problems that are not well suited for an approximation on a small n-dimensional linear subspace as created by \gls{POD}. Thus, a large reduced basis might be needed to obtain a good approximation quality of the \gls{FOM}, e.g., in high Reynolds number flow problems~\cite{Pichi.2022,Regazzoni.2024}. Although the \gls{INS} problem considered in this study is indeed nonlinear, linear methods are widely applied to it and can be observed to still provide a sufficiently good approximation~\cite{Khamlich.2023, VeigaPineiro.2025}.\\

Comparative studies of \gls{PODG} and \gls{PODI} have been conducted across diverse applications, and can be found for instance in~\cite{Czech.2022,Shin.2022,Karasozen.2022,Padula.2024,Schulze.2025,Bai.2021,Abbasi.2022,Halder.2022,Lee.2020,McQuarrie.2023,Hines.2023}. While most of them focus on accuracy and computational speed-up, a few studies extend the comparison to additional aspects. The authors in~\cite{Karasozen.2022} additionally investigate temporal extrapolation capabilities of \gls{ROM}s derived via Operator Inference and \gls{PODG} for the rotating thermal shallow water equation. Moreover, the authors in~\cite{Czech.2022} conduct an extended comparison for a structural dynamics problem. They show absolute computational cost savings depending on the number of online simulations. Furthermore, they analyze the extrapolation capabilities of the intrusive method and investigate, for the non-intrusive method, how training data size affects prediction errors for a fixed~\gls{RB} dimension.\\

For gas leakage events, \gls{ROM}s have been introduced in recent years for complex domains ranging from single offshore facilities and built-up areas~\cite{Kong.2026,Jing.2025,Yuan.2022,Song.2021,Shi.2018} to large-scale landscapes~\cite{Seo.2022,Xiang.2021,Na.2018}. In these works, various neural-network based non-intrusive \gls{ROM} approaches are used to predict the spatio-temporal concentration field, considering parametric dependencies like wind intensity, release rate or wind direction. Additional literature can be found in related research fields such as air pollution modeling. Like in the present work, air pollution transport can be modeled based on a wind field computation and a subsequent transport problem. However, contaminant does not originate from a point source but occurs rather as a distributed emission field. Publications introducing \gls{ROM}s related to urban air pollution studies can be found for instance in~\cite{Chen.2025,Khamlich.2023,Masoumi.2022}, featuring not only neural-network based approaches, but also intrusive methods.
The same holds for \gls{ROM}s concentrating on urban air flows. As shown in~\cite{Descartes.2025} and in references within \cite{Masoumi.2022}, the wind field can be identified as the dominant computational bottleneck, and \gls{ROM}s may be introduced to accelerate its prediction instead of directly predicting spatio-temporal concentration fields. Related to the sole prediction of urban air flows, further literature exists introducing surrogates for urban air flows as in the context of unmanned aerial vehicles~\cite{VeigaPineiro.2025}. 
Furthermore, resulting reduced models can be embedded in a framework for risk assessment as shown in \cite{Kong.2026} for an offshore hydrogen system, inside a digital twin, as shown in \cite{Xie.2025} for an indoor fire scenario, or in combination with evacuation route planning~\cite{Seo.2022}. Additionally, the authors in~\cite{Shi.2026} present a combination of a neural network based~\gls{ROM} with a large language model for decision-making. Besides real-time prediction of the wind field, the application of the \gls{ROM}s for uncertainty quantification is of utmost interest. To investigate the influence of different input parameters such as wind conditions or emission rates, sensitivity analyses were performed for instance in \cite{Chettouh.2021,Chettouh.2014,GarciaDiaz.2012}. The analyses follow a Monte Carlo scheme and evaluate results either point-wise for certain positions within the domain or at the plume center line. Studying whole field solutions, parameters and locations within the urban domain are identified in~\cite{GarciaSanchez.2017}, which are most sensitive to uncertainties, like inflow wind velocity and direction. Similarly, \cite{Rodriguez.2013} investigates the influence of spurious wind direction values on plume prediction in urban areas, demonstrating how such errors can lead to misalignment between the predicted and true concentration fields. These studies show the critical role of wind speed and direction as very important input parameters in contaminant dispersion modeling.\\

Based on the current research state described in the literature review presented above, this work presents a systematic, application-driven workflow. Here, special focus is laid on the model selection process in the first step, before continuing with constructing a corresponding~\gls{ROM}, and performing a Monte Carlo analysis for estimating the influence of potential wind measurement errors on the spatio-temporal concentration fields which can then be used to support evacuation planning or preparatory "what-if" scenarios. Hereby, the explicit focus on the model selection part extends aforementioned application-specific studies by considering relevant criteria such as accuracy, computational efficiency, data requirements, and extrapolation beyond the training parameter range. The latter two are less frequently covered by comparative literature, but are interesting additions for the considered case since they provide practical guidance for selecting the most suitable method based on available resources and the capability to extrapolate to higher wind velocities than originally considered during \gls{ROM} setup. While advanced non-intrusive neural-network based \gls{ROM}s have gained popularity, they often require extensive hyperparameter tuning, large training datasets, or specialized network architectures. In contrast, \gls{PODG} and \gls{PODI} require only minimal tuning with a straightforward dimensionality reduction that leverages the optimality of the \gls{POD} basis. This makes them ideal candidates as starting point for practical deployment. Therefore, this evaluation is applied, but not restricted to the two wide spread~\gls{PODG} and~\gls{PODI}. The methods are implemented for the incompressible Navier-Stokes equations with parametric dependence on wind intensity, which is the computational bottleneck for real-time simulation of airborne contaminant dispersion in complex environments. Based on the obtained insights, and accounting for the sensitivity of dispersion predictions on wind speed and direction, an extended non-intrusive ROM is constructed in the second step and its computational efficiency is subsequently leveraged to perform a computationally feasible Monte Carlo analysis. The analysis considers uncertain inflow conditions that could stem from inaccurate or noisy wind measurements and enables a probabilistic assessment of the concentration field, identifying areas with consistently high concentration values and areas where only a potential gas presence is indicated. This information is summarized by showing local exceedance probabilities, illustrating the probability of exceeding a certain concentration threshold, as it is of utmost importance for emergency response planning. Finally, we demonstrate the practical utility of the framework by embedding the ROM in a simple interactive dashboard, enabling real-time exploration of spatio-temporal concentration values. 
Summarized, the main contribution of the work is to present an application-driven workflow for benchmark-based gas dispersion analysis, which
\begin{itemize}
    \item starts from a detailed model selection process by providing a comparison between intrusive~\gls{PODG} and non-intrusive~\gls{PODI};
    \item accounts for potential wind measurement errors;
    \item and illustrates the extraction of valuable information for emergency response and preparatory studies in form of exceedance maps and an interactive dashboard.
\end{itemize}

The remainder of the article is structured as follows: Sec.~\ref{sec:background} reviews the mathematical and numerical modeling of the airborne contaminant transport problem as well as the considered \gls{MOR} techniques. Subsequently, Sec.~\ref{sec:num_prob} provides a comparison between \gls{PODG} and \gls{PODI} for the flow problem at hand, parametrized for different inflow wind velocities. This is followed by an extension of the \gls{PODI}-based \gls{ROM} to also cover different wind directions. In Sec.~\ref{sec:application}, the presented \gls{MOR} framework is applied to the crisis management scenario test-case described above and an interactive dashboard is shown, aiming at an easy accessibility and usability of the developed model. Conclusions are drawn in Sec.~\ref{sec:conclusion}, where possible future extensions of the study towards realistic three-dimensional turbulent flows are discussed.
\section{Mathematical Formulation of Full- and Reduced-Order Model}\label{sec:background}

\subsection{Full-Order Model}\label{sec:ins_fom}

To enable efficient benchmarking of the MOR methods, this study employs a parametrized two-dimensional laminar flow model governed by the steady-state \gls{INS} equations to evaluate the wind field and a subsequent advection-diffusion problem to analyze the contaminant transport. \\
For a two-dimensional bounded domain $\Omega \subseteq \R^2$, the \gls{INS} equations for the wind velocity field $\mbu$ and pressure value $\p$ read: 

\begin{equation}\label{eq:INS_strong}
    \begin{aligned}
        -\nu \nabla^2\mbu + (\mbu\cdot\nabla)\mbu + \nabla \p &= \mbnull \quad &\text{in} \quad &\Omega, \\
        \nabla \cdot \mbu &= 0 \quad &\text{in} \quad &\Omega, \\
        \mbu &= \mbg_D(\params) \quad &\text{on} \quad &\Gamma_D, \\
        \mbu &= \mbnull \quad &\text{on} \quad &\Gamma_0, \\
        \nu \frac{\partial \mbu}{\partial \mbn} - \p \mbn &= \mbnull \quad &\text{on} \quad &\Gamma_N,
        \end{aligned}
\end{equation}
where $\nu \in \R^+$ denotes the kinematic viscosity, $\Gamma_0$ no-slip boundaries at the buildings, and $\Gamma_N$ the Neumann outflow boundary with $\mbn$ being the outward-pointing normal to it. The effect of different inflow wind conditions is considered by the inhomogeneous Dirichlet boundary condition, which is parametrized by $\params \in \mathcal{D} \subset \R^d$, $d \in \{1,2\}$, depending on whether only the inflow wind velocity or additionally the wind direction varies, too. Thus, also the velocity and pressure are parameter-dependent quantities $\left(\mbu,\p\right) = \bigl(\mbu(\mbx,\params),\p(\mbx,\params)\bigl)$. The Reynolds number considered here accounts for the maximum velocity magnitude $u_{\mathrm{max}}$ so that $\Rey = u_{\mathrm{max}} \ l/\nu$, where $l$ is the characteristic length of the domain. 

To enable a numerical solution for the wind field, the Navier-Stokes equations in Eq.~\eqref{eq:INS_strong} are converted from the strong form to the weak form and are then discretized using Taylor-Hood finite elements, as described in~\cite{Ballarin.2015,Elman.2014}. This leads to a discretized system of equations for the velocity at every finite element node and in each spatial direction $\mbuh \in \R^{\fomdim}$ as well as the pressure value $\mbph \in \R^{\fomdimp}$. Here, $\fomdim$ and $\fomdimp$ represent the degrees of freedom related to the use of second-order Lagrange finite elements for the velocity and first-order Lagrange finite elements for the pressure. 
Following the derivations in \cite{Ballarin.2015}, the discretized form of the \gls{FOM} can be written as:
\begin{equation}\label{eq:ins_fom_discr}
    \begin{aligned}
    \begin{bmatrix}
    \mathbf{A}(\params) + \mathbf{C}(\mbu(\params));\params) & \mathbf{B}^\intercal(\params) \\
    \mathbf{B}(\params) & \mathbf{0}
    \end{bmatrix}
    \begin{bmatrix}
    \mbuh(\params) \\
    \mbph(\params)
    \end{bmatrix}
    =
    \begin{bmatrix}
    \mathbf{f}(\params) \\
    \mathbf{g}(\params)
    \end{bmatrix}
    \end{aligned}
\end{equation}
with $\mathbf{A},\mathbf{B}$, $\mathbf{C}$, $\mbf$ and $\mbg$ containing the discretized terms of the problem.
By means of the Newton-Raphson algorithm, implemented within the FEniCS framework~\cite{fenics}, the emerging nonlinear system of equations is eventually solved for pressure and velocity. The implementation is validated for well-known benchmark problems shown in~\cite{Elman.2014}.

Once the wind field has been computed by solving the \gls{INS} equations, the spatio-temporal gas concentration $c(\mathbf{x},t)$ is determined by solving the following advection-diffusion problem:

\begin{equation}\label{eq:ad_strong}
\begin{aligned}
  \frac{\partial c}{\partial t}-\kappa\nabla^2 c + \mbu\cdot\nabla c &= \parameterSource &\qquad&\text{in}\ \ \Omega \times \left(0,T\right),\\
  \kappa\nabla c \cdot \normal &= 0 &&\text{in} \left(\outflowBoundary \cup \characteristicBoundary\right) \times \left(0,T\right),\\ 
  c(\mathbf{x},t) &= 0 &&\text{in}\ \ \inflowBoundary \times \left(0,T\right),\\
  c(\mathbf{x},0) &= \parameterInitial &&\text{in}\ \ \Omega,
\end{aligned}
\end{equation}
featuring the diffusion coefficient $\kappa \in \R^+$ as well as the wind field $\mbu(\mbx,\params)$ calculated earlier. Following the definitions in \cite{Elman.2014}, $\outflowBoundary \subset \partial \Omega$ denotes the outflow boundary with $\mbu \cdot \mbn > 0$, $\characteristicBoundary \subset \partial \Omega$ the boundaries tangential to the flow such that $\mbu \cdot \mbn = 0$ and $\inflowBoundary \subset \partial \Omega$ the inflow boundary with $\mbu \cdot \mbn < 0$.  
Furthermore, the initial gas distribution is denoted by $\parameterInitial$, and $\parameterSource$ represents the source term of the advection-diffusion problem, cf.~\cite{Mattuschka.2026}. Hence, the formulation can describe instantaneous sources with $\parameterInitial$, as well as continuous gas leakages with $\parameterSource$. 

Selecting a first-order Lagrange finite element approach for the concentration and following \cite{Brooks.1982,Behr.2008,Danwitz.2023,KeyAD.2023,Mattuschka.2026}, a \gls{SUPG}-stabilized weak form for the boundary value problem can be derived. To address the time-dependent nature of the problem, an implicit backward Euler time-stepping algorithm is applied as described in~\cite[Eq.~(10.25)]{Elman.2014}. Altogether, space and time discretization ultimately lead to a linear system of equations to be solved for the unknown gas concentration field. 

\subsection{Reduced INS Problem}
A \gls{ROM} is constructed for the \gls{INS} problem in Eq.~\eqref{eq:INS_strong} using two methods: the intrusive \gls{PODG} approach combined with the discrete empirical interpolation method and the non-intrusive counterpart, \gls{POD} with interpolation. The two approaches first construct a reduced basis by applying \gls{POD} to the snapshot solutions obtained from the full-order model. The goal of both methods is to efficiently map unseen parameter values to their corresponding approximate solutions of Eq.~\eqref{eq:INS_strong}. 

\subsubsection*{Proper orthogonal decomposition}\label{sec:pod}
To construct a reduced basis via \gls{POD}, snapshot solutions of the high-fidelity model are computed in a first step for specific training parameters $\paramstrain \in \trainset = \{\params_1,\hdots,\params_{\nsnap}\}$, which are determined based on a-priori knowledge of the system. These $\nsnap$ snapshots are collected as columns of the snapshot matrix $\snap = [\s_1,\hdots, \s_{\nsnap}]$, where $\s_i = \s(\params_i)$ is the \gls{FOM} solution corresponding to the $i^{\mathrm{th}}$ parameter value in $\trainset$. For the specific application and the associated focus on the velocity, the following is shown for velocity snapshots, so that $\mathbf{s}_i = \mbuh_i$ and $\snap  \in \R^{\fomdim \times \nsnap}$. However, similar relations hold for pressure snapshots, which would result in different dimensions than those presented here. Ultimately, the aim of \gls{POD} is to find a low-dimensional linear subspace $\subspace = \spn\{\projmatrix_1,\hdots,\projmatrix_{\romdim}\}$, spanned by the column-orthogonal \gls{POD} basis matrix $\projmatrix = [\projmatrix_1,\hdots,\projmatrix_{\romdim}] \in \R^{\fomdim \times \romdim}$ such that the error 

\begin{equation}
    \delta_{\mathrm{rb}} = \sum_{k=1}^{\nsnap} \big\|\s_k - \projmatrix\projmatrix^\intercal\s_k\big\|^2_2
\end{equation}
between the snapshot matrix and its orthogonal projection onto $\subspace$ is minimized \cite{BennerVol1.2021}. Such a reduced basis can be found via a singular value decomposition (\gls{SVD}) of the snapshot matrix 
\begin{equation}
    \snap = \mathbf{U}\boldsymbol{\Sigma}\mathbf{Z}^\intercal,
\end{equation} 
with $\mathbf{U} = [\boldsymbol{\Phi}_1,\hdots,\boldsymbol{\Phi}_{\fomdim}] \in \R^{\fomdim \times \fomdim}$ and $\mathbf{Z} = [\boldsymbol{\Psi}_1,\hdots,\boldsymbol{\Psi}_{\nsnap}] \in \R^{\nsnap \times \nsnap}$ being orthogonal matrices, as well as $\boldsymbol{\Sigma} = \diag\left({\sigma}_{1},\hdots,{\sigma}_{\rank}\right) \in \R^{\fomdim \times \nsnap}$ the eigenvalue matrix, where $\rank \leq \mathrm{min}(\fomdim,\nsnap)$ denotes the rank of the snapshot matrix and $\sigma_1 \geq \sigma_2 \geq \hdots \geq \sigma_{\rank}$~\cite{Quarteroni.2015}. From the \gls{SVD}, it follows that
\begin{equation} \label{eq:svd_evp}
    \snap \mathbf{\Psi}_i = \sigma_i\mathbf{\Phi}_i \quad \text{and} \quad \snap^\intercal\mathbf{\Phi}_i = \sigma_i\mathbf{\Psi}_i, \quad i = 1,\hdots,\rank.
\end{equation}
Inserting the two expressions into each other leads to the equivalent formulation
\begin{equation}\label{eq:corr_evp}
    \snap^\intercal \snap \mathbf{\Psi}_i = \sigma_i^2 \mathbf{\Psi}_i \quad \text{and} \quad  \snap \snap^\intercal \mathbf{\Phi}_i = \sigma_i^2 \mathbf{\Phi}_i, \quad i = 1,\hdots,\rank.
\end{equation}
The eigenvalue problems in Eq.~\eqref{eq:corr_evp} enable an efficient determination of the basis functions without having to solve the full singular value decomposition. Furthermore, since the number of degrees of freedom commonly vastly exceeds the number of snapshots, it is computationally preferable to solve the first eigenvalue problem in Eq.~\eqref{eq:corr_evp} featuring the symmetric positive-definite correlation matrix $\mathbf{C} = \snap^\intercal \snap \in \R^{\nsnap\times\nsnap}$, or $C_{ij}=(s_i,s_j)_{L^2}$, respectively \cite{Quarteroni.2015,Benner.2020,Khamlich.2023}. Although this is beneficial in terms of computational resources, the authors in \cite{Ballarin.2015} warn that constructing the reduced basis via the correlation matrix may lead to poorer conditioning. Finally, the desired basis functions $\mathbf{\Phi}_i$ can be obtained by means of the eigenvectors of the correlation matrix $\mathbf{\Psi}_i$ by reformulating the relation in Eq.~\eqref{eq:svd_evp} as 
\begin{equation}
    \mathbf{\Phi}_i = \frac{1}{\sigma}_i \snap \mathbf{\Psi}_i, \quad i = 1,\hdots,\romdim.
\end{equation}
In the following, a truncation is performed to retain only information associated with the largest $\romdim \ll \fomdim$ eigenvalues of $\snap \snap^\intercal$, along with their corresponding eigenvectors. These eigenvectors span the desired reduced subspace $\subspace$, and are used to construct the basis matrix $\mathbf{V} = [\mathbf{\Phi}_1, \hdots,\mathbf{\Phi}_{\romdim}] \in \R^{\fomdim \times \romdim}$~\cite{BennerVol1.2021}. Repeating the procedure for the pressure, a subspace $\mathcal{P}$ is found, spanned by the columns of $\mathbf{P} = [\mathbf{\Upsilon}_1, \hdots,\mathbf{\Upsilon}_{\romdimp}] \in \R^{\fomdimp \times \romdimp}$. With the parameter-independent eigenvectors -- also called \gls{POD} modes -- quantities can eventually be approximated by a linear combination of \gls{POD} modes as well as parameter-dependent coefficients
\begin{equation}\label{eq:POD_approx}
    \mbuh(\params) \approx \sum_{i=1}^{\romdim} a_i(\params) \mathbf{\Phi}_i = \projmatrix \mathbf{a}(\params) \quad \text{and} \quad \mbph(\params) \approx \sum_{i=1}^{\romdimp} b_i(\params) \mathbf{\Upsilon}_i = \projmatrixp \mathbf{b}(\params). 
\end{equation} 
While the first formulation of the approximation above emphasizes the \gls{POD} modes and coefficients, it is often equivalently expressed as a matrix vector multiplication of the basis matrices $\projmatrix$ and $\projmatrixp$, and the \gls{POD} coefficients $\mathbf{a}(\params) \in \R^{\romdim}$ and $\mathbf{b}(\params) \in \R^{\romdimp}$. The latter are referred to as the reduced velocity $\mbur(\params) \in \R^{\romdim}$ and pressure $\mbpr(\params) \in \R^{\romdimp}$, i.e., $\mbur(\params) = \mathbf{a}(\params)$ and $\mbpr(\params) =\mathbf{b}(\params)$.

Due to the fact that the error introduced by \gls{POD} is directly proportional to the magnitude of the neglected eigenvalues \cite{Benner.2020}, a fast decay of the eigenvalues is crucial for a good approximation quality. A measure of the retained energy can be given as 
\begin{equation}
    E_{\mathrm{rb}} = \frac{\sum_{i=1}^{\romdim} \sigma_i^{2}}{\sum_{k=1}^{\rank} \sigma_k^{2}} \quad \in [0,1].
\end{equation}
Therefore, slowly decaying eigenvalues, and with that a slowly decaying Kolmogorov n-width, indicate that the system at hand is not easily reducible by introducing a low-dimensional subspace as derived here. Once the reduced basis is established, the full-order solutions for any given parameter value can be approximated by solving either a projected system of equations (\gls{PODG}) or an interpolation/regression task (\gls{PODI}) to determine the \gls{POD} coefficients, i.e., the reduced velocity and pressure.

\subsubsection*{POD with Galerkin projection}\label{sec:PODGalerkin_background}
As described in the previous paragraph, \gls{PODG} utilizes the computed reduced basis functions, also known as the \gls{POD} modes, for velocity and pressure, to project the system of equations of the \gls{FOM} onto the subspaces they span. By further replacing the full-order quantities with the approximation as shown in  Eq.~\eqref{eq:POD_approx}, the system of equations to be solved online reduces to the size of the reduced basis for velocity and pressure.
By performing a Galerkin projection on the discretized system of equations in Eq.~\eqref{eq:ins_fom_discr}, the residual that results from replacing $\mbuh$ with the approximation in Eq.~\eqref{eq:POD_approx} is orthogonal to the subspace spanned by the columns of $\projmatrix$ and $\projmatrixp$, i.e., the \gls{POD} modes \cite{Ballarin.2015}. This can be written as
\begin{equation}
\begin{aligned}
\mathbf{\mathcal{R}} &= 
\begin{bmatrix}
    \mathbf{A}(\params) + \mathbf{C}(\projmatrix\mbur(\params));\params) & \mathbf{B}^\intercal(\params) \\
    \mathbf{B}(\params) & \mathbf{0}
    \end{bmatrix}
    \begin{bmatrix}
    \projmatrix\mbur(\params) \\
    \mathbf{P}\hat{\mathbf{p}}(\params)
    \end{bmatrix}
    - \begin{bmatrix}
    \mathbf{f}(\params) \\
    \mathbf{g}(\params)
    \end{bmatrix} \\
    \quad
    &\Rightarrow
    \quad
    \begin{bmatrix}
    \projmatrix^\intercal & \mathbf{0} \\
    \mathbf{0} & \mathbf{P}^\intercal
    \end{bmatrix} 
    \mathbf{\mathcal{R}} = \mathbf{0}.
\end{aligned}
\end{equation}
With that, the reduced set of equations can be expressed as \cite{Ballarin.2015}:
\begin{equation}\label{eq:POD_Galerkin_ROM}
    \begin{bmatrix}
    \redA(\params) + \redC(\mbur(\params));\params) & \redB^\intercal(\params) \\
    \redB(\params) & \mathbf{0}
    \end{bmatrix}
    \begin{bmatrix}
    \mbur(\params) \\
    \mbpr(\params)
    \end{bmatrix}
    =
    \begin{bmatrix}
    \redf(\params) \\
    \redg(\params)
    \end{bmatrix}
\end{equation}
with
\begin{equation*}
    \begin{aligned}
    &\redA(\params) = \projmatrix^\intercal\mathbf{A}(\params)\projmatrix, \ \redC = \projmatrix^\intercal\mathbf{C}(\projmatrix\mbur(\params);\params)\projmatrix, \ \redf(\params) = \projmatrix^\intercal\mbf(\params),\\
    &\redB(\params) = \projmatrixp^\intercal\mathbf{B}(\params)\projmatrixp, \ \redg(\params) = \projmatrix^\intercal\mbg(\params).
    \end{aligned}
\end{equation*}
Although this description is based on formulating the projection as matrix multiplication to the discretized operators stemming from the weak form of Eq.~\eqref{eq:INS_strong}, other formulations can be found in literature, which rely on the superposition of \gls{POD} modes scaled by their coefficients as shown in \cite{Khamlich.2023,Hijazi.2023,Lorenzi.2016}. However, the result is eventually the same -- a nonlinear system of equations for the \gls{POD} coefficients of Eq.~\eqref{eq:POD_approx} for any new parameter value.
Since the implementation of the steps above requires access to the solver, the \gls{PODG} approach is an intrusive approach.

To enable an efficient online-offline decomposition, parameter-dependent terms in the equations above are affinely decomposed into a parameter-dependent factor and a parameter-independent matrix or vector \cite[Sec.~2.3]{Ballarin.2015}. For low-order polynomial nonlinearities and affine parameter dependencies, the respective system of equations can be expressed such that an efficient online-offline decomposition can be achieved~\cite{Quarteroni.2015,Ballarin.2015}. However, as suggested in~\cite{Ballarin.2015}, the application of the discrete empirical interpolation method (\gls{DEIM})~\cite{Chat.2010} can help reducing required storage costs by approximating the nonlinear term in Eq.~\eqref{eq:POD_Galerkin_ROM} and its derivative, needed in the context of a nonlinear solution scheme.

\subsubsection*{Discrete Empirical Interpolation Method}\label{sec:DEIM}
The key motivation of \gls{DEIM}, introduced in \cite{Chat.2010}, is to ensure an efficient offline-online decomposition for nonaffine parameter dependencies or nonlinear problems with high-dimensional or non-polynomial nonlinearities. For the presented Navier-Stokes equations, \gls{DEIM} can be applied to reduce storage costs~\cite{Ballarin.2015} for the convective term and its derivative as needed for the Jacobian within a Newton scheme. Focusing on the system in Eq.~\eqref{eq:POD_Galerkin_ROM}, the algorithm can be illustrated by splitting the solution-dependent terms of the nonlinear problem into a linear part $\mbL(\params)\mbu(\params)$, and a nonlinear part $\mbN(\mbu(\params))$~\cite{Xiao.2014}. With $\mbL \in \R^{\fomdim\times \fomdim}$, $\mbL(\params)\mbu(\params)$ and $\mbN(\mbu(\params))$ describe vector-valued functions. For the Navier-Stokes problem, the nonlinear part reads as $\mbN(\mbu(\params)) = \mathbf{C}(\mbu(\params);\params)\mbu(\params)$. After applying the POD-Galerkin dimensionality reduction, this results in:
\begin{equation}
  \hat{\mbN}(\mbu(\params))=\projmatrix^{\intercal}\mbN(\projmatrix\mbur(\params)) = \underbrace{\projmatrix^{\intercal}}_{\romdim \times \fomdim}\underbrace{\mathbf{C}(\projmatrix\mbur(\params);\params)\projmatrix\mbur(\params)}_{\fomdim \times 1}.
\end{equation}
Hence, the nonlinear term has to be evaluated for every $\params$ requiring \gls{FOM} operator evaluations during the online stage. To circumvent this, the affine approximation of the nonlinear functions $\mbN(\projmatrix\mbur(\params))$, for the sake of brevity represented by $\mbc(\params)$, is found via projection onto a low-dimensional subspace spanned by a $M \ll \fomdim$ dimensional basis $\mbQ = \left[\mbrho_1,\hdots,\mbrho_M\right] \in \R^{\fomdim \times M}$: 
\begin{equation}\label{eq:deim_ansatz}
    c(\params) \approx c_M(\params) = \mbQ \mbgamma(\params).
\end{equation}
Here, $\mbgamma(\params) \in \R^{M}$ denotes the coefficients of the approximation.
The \gls{DEIM} offline stage begins with the basis generation. For this, POD is applied on snapshots of the nonlinear expression which can be derived from wind field evaluations computed for the POD dimensionality reduction in Sec.~\ref{sec:pod}. Thereafter, $M$ indices $\mathcal{J} \subset \{1,\hdots,\fomdim\}$, $|\mathcal{J}| = M$, are chosen corresponding to $M$ rows of the overdetermined system in Eq.~\eqref{eq:deim_ansatz}~\cite{Chat.2010, Quarteroni.2014}. They are determined via a greedy algorithm, that iteratively adds indices to the set $\mathcal{J}$ for which the current \gls{DEIM} approximation is most inaccurate. This can be done without explicitly determining the unknown coefficients $\mbgamma(\params)$ by directly inserting the interpolation constraints. These ensure an exact approximation at the (current) interpolation points in $\mathcal{J}$. The constraints read as:
\begin{equation}
    \mbQ_{\mathcal{J}}\mbgamma(\params) = \mbc_{\mathcal{J}(\params)},
\end{equation}
where $\mbQ_{\mathcal{J}} \in \R^{|\mathcal{J}|\times |\mathcal{J}|}$ contains the rows of $\mbQ$ corresponding to indices in $\mathcal{J}$ and, accordingly, $\mbc_{\mathcal{J}(\params)} \in \R^{|\mathcal{J}| \times 1}$ refers to entries of $\mbc$ at the indices in $\mathcal{J}$. The algorithm stops when the \gls{DEIM} approximation reaches a certain accuracy~\cite[Algorithm~10.3]{Quarteroni.2014}. For any new parameter value, the coefficients $\mbgamma(\params)$ can then be computed online with the interpolation constraints, so that the resulting expression can be written as:
\begin{equation}
    \mbc_M(\params) = \mbQ\mbQ_{\mathcal{J}}^{-1}\mbc_{\mathcal{J}}(\params).
\end{equation}
Reintroducing the Navier-Stokes term instead of $\mbc_{\mathcal{J}}$, a formulation is found to approximate the reduced nonlinear term as:
\begin{equation}
\begin{aligned}
\hat{\mbN}(\mbu(\params)) \approx \projmatrix^{\intercal}\mbc_M(\params) &= \projmatrix^{\intercal} \mbQ\mbQ_{\mathcal{J}}^{-1}\mbN(\projmatrix_{\mathcal{J}}\mbur(\params))\\ &= \underbrace{\projmatrix^{\intercal} \mbQ\mbQ_{\mathcal{J}}^{-1}}_{\text{pre-computable}, \ \romdim \times M} \underbrace{\mathbf{C}(\projmatrix_{\mathcal{J}}\mbur(\params);\params)\projmatrix_{\mathcal{J}}\mbur(\params).}_{M \times 1}
\end{aligned}
\end{equation}
The operational costs during the online phase to evaluate the aforementioned parametric term now depend on $M$~\cite[Chap.~10.3]{Quarteroni.2015}, whereas operations depending on the full \gls{FOM}-size can be pre-computed during the offline stage due to the independence on the parameter space. It has to be noted, however, that when considering a finite element implementation, neighboring nodes need to be included for computing the DEIM enhanced nonlinear term. The union of DEIM interpolation points and their neighboring nodes can be denoted as reduced mesh, so that the actual computational costs depend on the reduced mesh rather than $M$ alone. The DEIM-enhanced nonlinear term is then obtained by assembling the correspondent finite element form over the reduced mesh, where the integrand is evaluated on the quadrature points of the reduced mesh cells~\cite{Cicci.2022}. Moreover, the reader should note, that the respective ROM remains nonlinear, yet computational inefficiencies due to the nonlinearity can be avoided via the \gls{DEIM} approximation. For details on the application directly to the residual and Jacobian, as well as an extended description of the algorithm, the reader is referred to \cite{Chat.2010, Xiao.2014, Quarteroni.2015,Cicci.2022}.

\subsubsection*{POD with interpolation}\label{sec:PODI}
Although the initial steps of \gls{PODI} -- computing snapshots and constructing a reduced basis via \gls{POD} -- are identical to those in \gls{PODG}, \gls{PODI} operates non-intrusively by using interpolation or regression methods instead of a Galerkin projection for mapping unknown input parameter values to \gls{POD} coefficients. For this purpose, \gls{PODI} relies solely on the snapshot data. Hence, rather than projecting the system of equations, only the snapshot data is projected: 

\begin{equation}\label{eq:snap_proj_podi}
     \snap \approx \projmatrix \hat{\snap}  \quad \text{or} \quad \mathbf{s}(\params_k) \approx \sum_{i=1}^{\romdim} \hat{s}_{i}(\params_k) \mathbf{\Phi}_{i}
\end{equation}
with $\hat{\snap} = \projmatrix^\intercal \snap$ or $\hat{s}_{i}(\params_k) = \left(\mathbf{\Phi}_{i},\mathbf{s}(\params_k)\right)_{L^2}$ and $k=1,\hdots,\nsnap$, respectively. Here, $\hat{\snap} \in \R^{\romdim \times \nsnap}$ and $\hat{s}_{i}(\params_k) \in \R$ denote the \gls{POD} coefficients in Eq.~\eqref{eq:POD_approx} for an approximation of the \gls{FOM} solutions collected inside the snapshot matrix. This set of known \gls{POD} coefficients is now used to set up an approximation of the mapping between parameter values and \gls{POD} coefficients $\boldsymbol{f}: \mathcal{D} \ \xrightarrow{} \R^{\romdim}, \params \ \mapsto \mathbf{a}$ by means of an interpolation or regression scheme $\boldsymbol{\pi}: \mathcal{D} \ \xrightarrow{} \R^{\romdim}$. 
To enable such a mapping, the set of training parameter values $\trainset$ is needed as isolated input and therefore has to be stored separately with the $i^{\mathrm{th}}$ parameter value corresponding to the $i^{\mathrm{th}}$ snapshot.
The interpolation or regression can then be performed with various algorithms such as radial basis function (\gls{RBF}) interpolation, Gaussian process regression or artificial neural networks \cite{Yu.2019,Berzins.2020}. Since it showed superior results for the conducted numerical experiments, the \gls{RBF} interpolation is used here, so that $\boldsymbol{\pi}(\params)$ reads as:
\begin{equation}\label{eq:rbf_interpol}
    \boldsymbol{\pi}(\params) = \sum_{k=1}^{\nsnap}\mathbf{w}_{k}\varphi(\|\params - \params_k\|), \quad \text{where} \quad \pi_{i}(\params_k) = \hat{s}_{i}(\params_k), \ i = 1,\hdots,\romdim.
\end{equation}
Here, the scalar-valued radial basis function $\varphi(d)$ corresponds to the Euclidean distance $d$ between a chosen sample point $\params$ and a so-called center, in this case the parameter values $\params_k$ that were considered when taking the snapshots. There exist many different radial basis functions in literature, and one example -- which is employed going forward -- is the thin-plate spline $\varphi(d) = d^2\ \log(d)$. Additionally, $\mathbf{w}_{k} \in \R^{\romdim}$ denotes the $k^{\mathrm{th}}$ weighting vector that is set up offline such that the \gls{POD} coefficients $\hat{s}_{i}(\params_k)$ stemming from the snapshots can be reproduced exactly \cite{Berzins.2020,Xiao.2015}.\\
Following the offline phase consisting of the projection of snapshot data and computing the weighting vectors, Eq.~\eqref{eq:rbf_interpol} can be evaluated for new parameter values, delivering interpolated \gls{POD} coefficients to eventually approximate the \gls{FOM} solution. Thus, the approximation in Eq.~\eqref{eq:POD_approx} can be written as
\begin{equation}
    \mbuh(\params) \approx \sum_{i=1}^{\romdim} \pi_{i}(\params) \mathbf{\Phi}_{i} = \projmatrix \boldsymbol{\pi}(\params),
\end{equation}
where \gls{POD} coefficients are approximated by the interpolation scheme for any new parameter value $\params$, so that $a_{i}(\params) \approx \pi_{i}(\params)$.

Although the procedure above is written for setting up one \gls{ROM} which represents the full nodal velocity vector $\mbuh_{j} = \left[u^h_x, u^h_y\right]_{j}$, it can be beneficial to set up two separate \gls{RB}s -- one for each of the two spatial velocity components. In this case, the nodal velocity vector for each snapshot is split up and two separate matrices are constructed on which, subsequently, POD is performed to gain two \gls{RB}s. Also the following setup of the RBF interpolation follows this split, leading to two decoupled \gls{PODI} approximations for x- and y-component of the velocity.

\section{Numerical Examples}\label{sec:num_prob}
\subsection{Comparing PODG and PODI for INS with varying inflow wind velocities}\label{sec:comparison_PODG_PODI}

\begin{figure}
    \centering
    \begin{subfigure}[b]{0.45\textwidth}
    \includegraphics[scale=0.7]{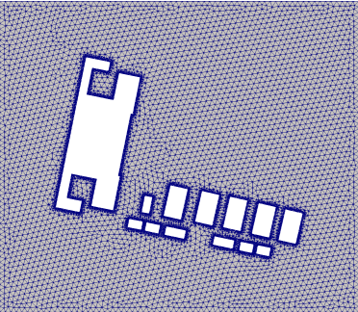}
    \caption{ }
    \label{fig:mesh_geom1}
    \end{subfigure}
    \begin{subfigure}[b]{0.45\textwidth}
    \hspace*{1.45em}
    \includegraphics[width=\textwidth]{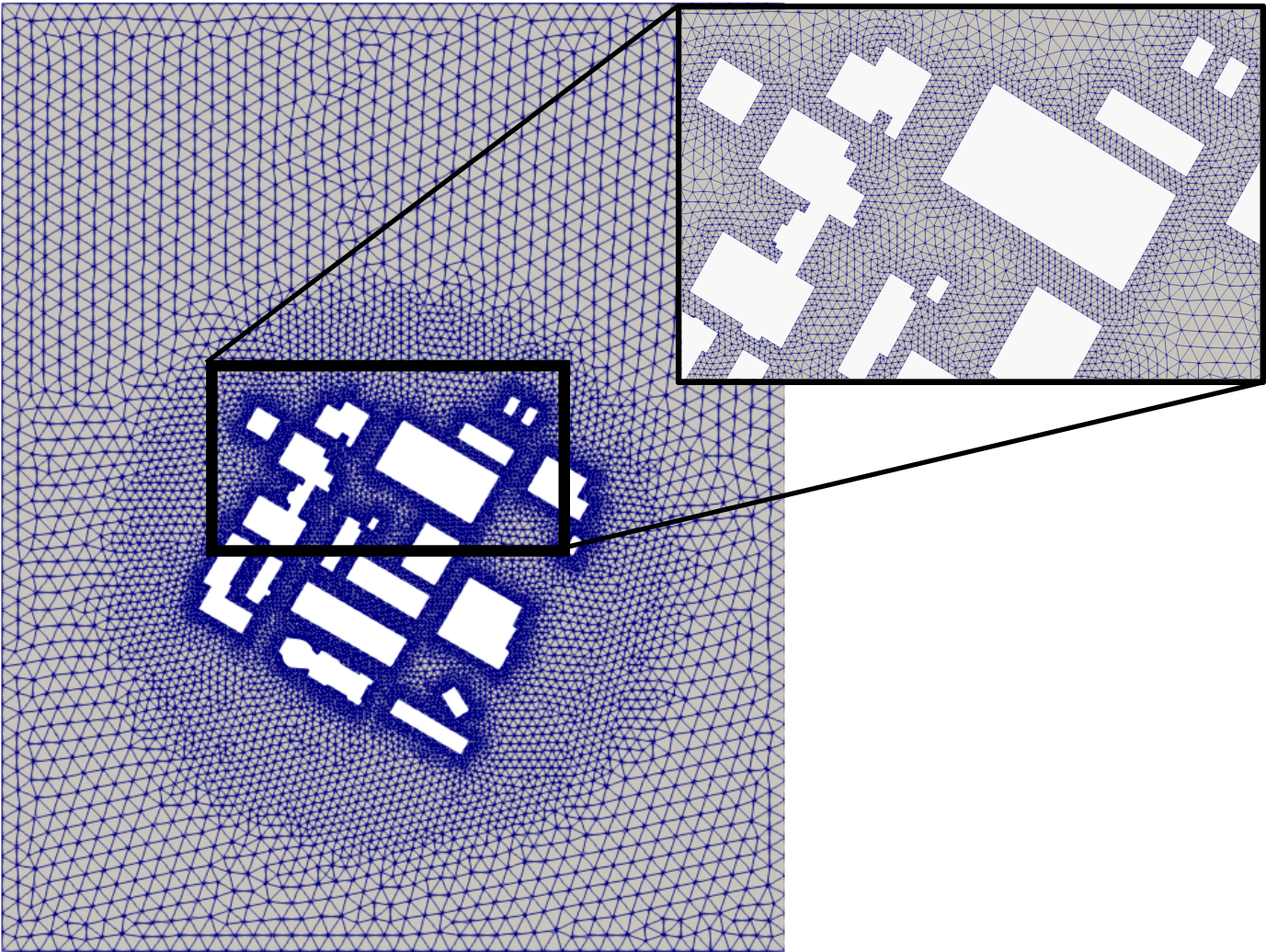}
    \caption{ }
    \label{fig:mesh_geom2}
    \end{subfigure}
    \caption{Meshed computational domains. Geometry 1 (a) features a sparse set of buildings, while Geometry 2 (b) represents a denser development. Both domains are based on building footprints derived from real, geo-referenced data, cf.~\cite{Bonari.2024}.}
    \label{fig:meshes}
\end{figure}

Due to the downstream relevance in the discussed applications (Sec.~\ref{sec:application}) we focus on evaluating the \gls{ROM} for the wind field and omit the discussion of the pressure field results. To compare the performance of \gls{PODG} with its non-intrusive counterpart \gls{PODI}, the wind field is reduced for two different real-world geometries of varying complexity, referred to as Geometry 1 and Geometry 2. The computational domains are derived as shown in \cite{Bonari.2024}, and their mesh discretization can be seen in Fig.~\ref{fig:meshes}. The setup follows Sec.~\ref{sec:ins_fom}, with additional details provided in Table \ref{tab:comparison_foms}. Due to the constant wind direction, the characteristic length used in the Reynolds number is chosen to be the inflow boundary. 

\begin{table}
\centering
\caption{Parameters of the full-order models considered for the comparison between \gls{PODG} and \gls{PODI}.}
\begin{tabular}{c c c}\toprule
\multirow{2}*{\textbf{Parameter}} &\multicolumn{2}{c}{\textbf{Numerical Examples}} \\
\cline{2-3}
 & \textbf{\textit{Geometry 1}} & \textbf{\textit{Geometry 2}} \\
\midrule
Degrees of freedom & 64,122 & 71,742 \\ 
\hline
Reynolds number & $\Rey \in \left[5.5,205\right]$ & $\Rey \in \left[10,380\right]$\\ 
\hline
Inflow wind velocity ($\si{\meter\per\second}$) &  $\wint \in \left[0.5,20\right]$ & $\wint \in \left[0.5,20\right]$ \\
\hline
Inflow wind direction & South & South \\
\bottomrule
\end{tabular}
\label{tab:comparison_foms}
\end{table}

\begin{figure}
    \centering
    \begin{subfigure}[b]{0.47\textwidth}
    \hspace*{-1.25em}
    \includegraphics[width=\textwidth, keepaspectratio]{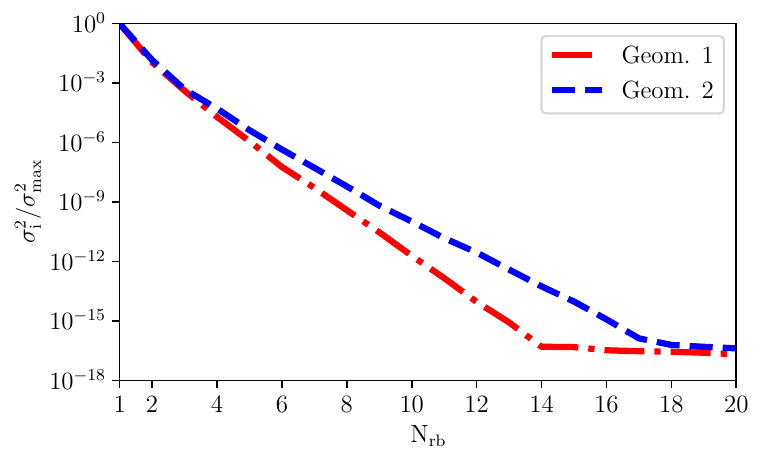}
    \caption{}
    \label{fig:eigvals_ins_only}
    \end{subfigure}
    \begin{subfigure}[b]{0.45\textwidth}
    \hspace*{-1.25em}
    \includegraphics[width=\textwidth, keepaspectratio]{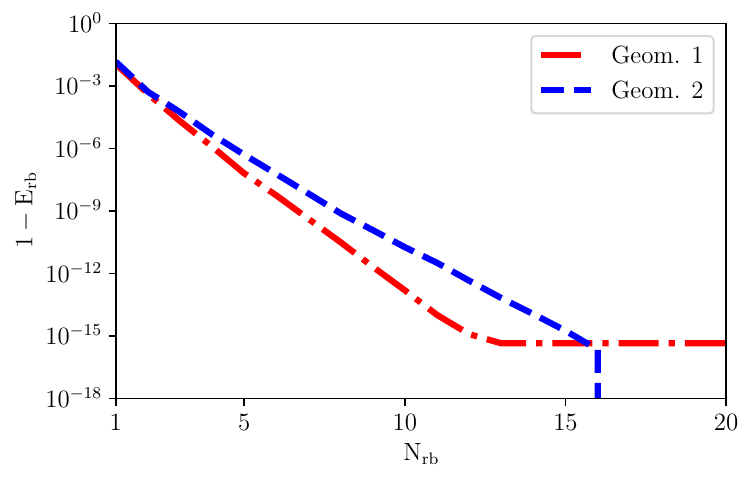}
    \caption{}
    \label{fig:retained_energy_ins_only}
    \end{subfigure}
    \caption{Normalized eigenvalues of the correlation matrix for snapshot solutions of the wind velocity (a) and the retained energy (b) as a function of the reduced basis size. Note that the observed drop for Geometry 2 with $\romdim > 16$ in (b) occurs because $E_\mathrm{rb} = 1$ for these \gls{RB} sizes.}
    \label{fig:redbarkeit_ins_only}
\end{figure}

To assess the reducibility of the problem, the normalized eigenvalues $\sigma_i^{2}/\sigma_\mathrm{max}^{2}$ of the correlation matrix are analyzed and the results are shown in Fig.~\ref{fig:eigvals_ins_only}. It can be observed, that for both geometries, the respective eigenvalues decay quickly, indicating that both problems are well-suited for finding an appropriate approximation within a low-dimensional subspace. In Fig.~\ref{fig:retained_energy_ins_only}, the retained energy is also shown as a first indicator of the necessary size of the reduced basis. To ensure a fair comparison of the two reduction methods, identical sets of snapshot data are used for both methods in terms of training and testing.

\subsubsection*{PODG}
Since the \gls{INS} problem is nonlinear, the \gls{PODG} approach is combined with the discrete empirical interpolation method~\cite{Chat.2010} for the convective term as described in the previous section. The respective parameters for the \gls{ROM} can be found in Table~\ref{tab:Parameters_MOR}. Note that training and testing parameter values are sampled equidistantly throughout the parameter space to ensure uniform coverage of the parameter space. Yet, other sampling strategies can be found, for instance in~\cite{Quarteroni.2015}. The implementation used in this study is based on the open-source tool~\emph{RBniCS}~\cite{rbnics,Rozza.2024}. 

\begin{table}
\caption{Parameters of the \gls{PODG} and \gls{PODI} \gls{ROM}, parametrized by the inflow wind velocity. Note that \gls{DEIM} quantities are not applicable to the \gls{PODI} approach.}
\begin{center}
\begin{tabular}{c c c c}\toprule

\multicolumn{2}{c}{\multirow{2}*{\textbf{Parameter}}} &\multicolumn{2}{c}{\textbf{Numerical Examples}} \\
\cline{3-4}
\multicolumn{2}{c}{ }& \textbf{\textit{Geometry 1}} & \textbf{\textit{Geometry 2}} \\
\midrule
Subspace & \gls{POD} ($\romdim$)& \multicolumn{2}{c}{$\romdim \in [1,20]$} \\ 
\cline{2-4}
dimension & \gls{DEIM} ($\deimdim$) & \multicolumn{2}{c}{20} \\ \hline
Snapshots & (\gls{POD}, \gls{DEIM}) & \multicolumn{2}{c}{50} \\ \hline
Test set size & (\gls{POD}, \gls{DEIM}) & \multicolumn{2}{c}{20} \\ \bottomrule
\end{tabular}
\label{tab:Parameters_MOR}
\end{center}
\end{table}

\subsubsection*{PODI}
For the reduction by means of \gls{PODI}, a \gls{RB} is created by \gls{POD} for each velocity component, and the mapping between input parameters and \gls{POD} coefficients is realized via \gls{RBF} interpolation with a thin-plate spline kernel as shown in Sec.~\ref{sec:PODI}. Other parameters remain constant compared to the \gls{PODG} approach as shown in Table~\ref{tab:Parameters_MOR}, except for those related to \gls{DEIM}, which is not required when applying \gls{PODI}. The implementation is based on the open-source toolbox~\emph{EZyRB}~\cite{Demo.2018}. It should be noted that, unlike~\emph{RBniCS}, no supremizer stabilization or lifting procedure is implemented in~\emph{EZyRB}. However, during the course of the evaluations below, neither instabilities nor issues with boundary conditions could be observed. Moreover, the decoupled treatment of the velocity components was implemented for the following numerical studies as it led to slightly better predictions than with one \gls{RB} containing information on spatial velocity components. However, below reported timings and speed-up of the \gls{PODI} approach always account for setup and evaluation for both velocity components.

\subsubsection*{Results}
\begin{figure}
    \centering
    \begin{subfigure}[b]{0.47\textwidth}
    \hspace*{-1.25em}
    \includegraphics[width=\textwidth, keepaspectratio]{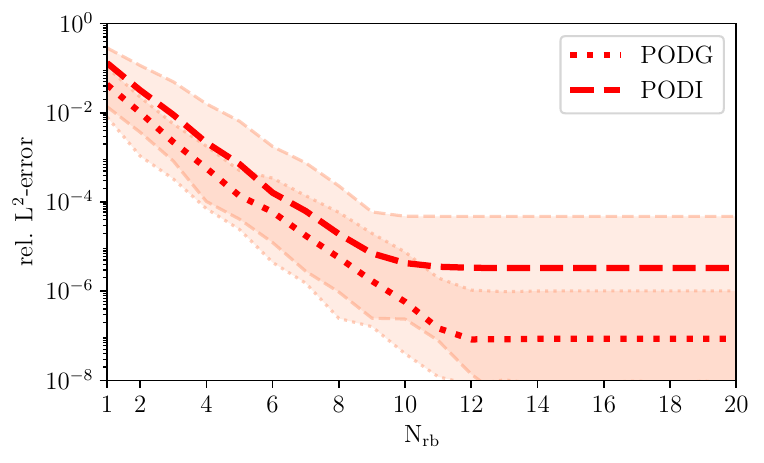}
    \caption{ }
    \label{fig:mean_rel_l2_campus_ins_only}
    \end{subfigure}
    \begin{subfigure}[b]{0.47\textwidth}
    \hspace*{-1.25em}
    \includegraphics[width=\textwidth, keepaspectratio]{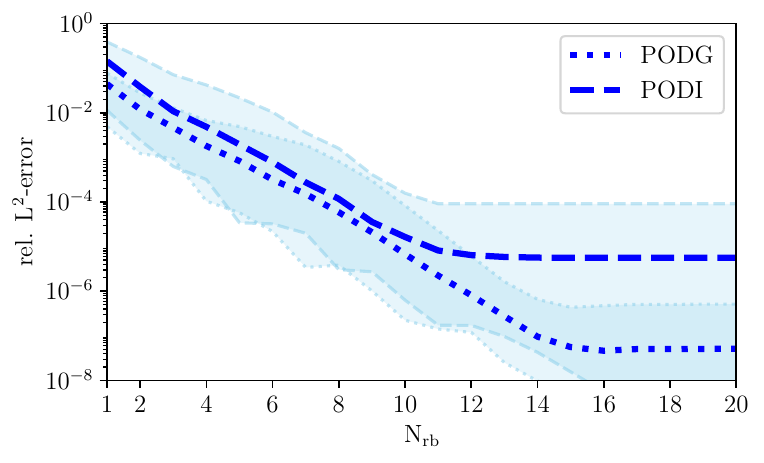}
    \caption{ }
    \label{fig:mean_rel_l2_henkel_ins_only}
    \end{subfigure}
    
    \begin{subfigure}[b]{0.47\textwidth}
    \hspace*{-1.25em}
    \includegraphics[width=\textwidth, keepaspectratio]{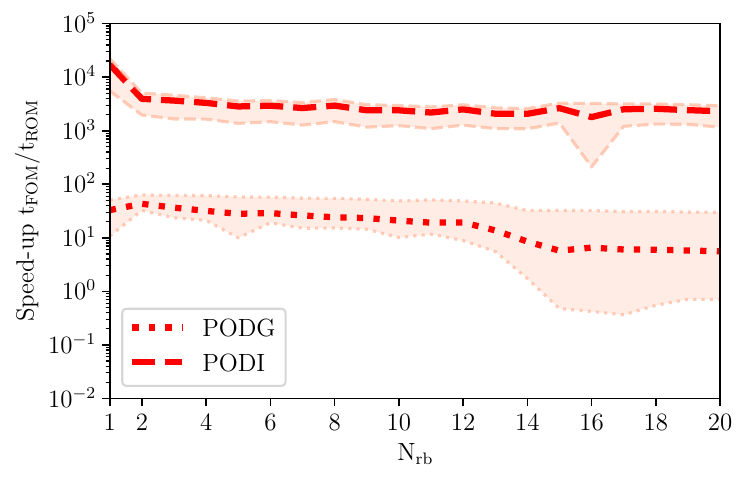}
    \caption{ }
    \label{fig:speedup_campus_ins_only}
    \end{subfigure}
    \begin{subfigure}[b]{0.47\textwidth}
    \hspace*{-1.25em}
    \includegraphics[width=\textwidth, keepaspectratio]{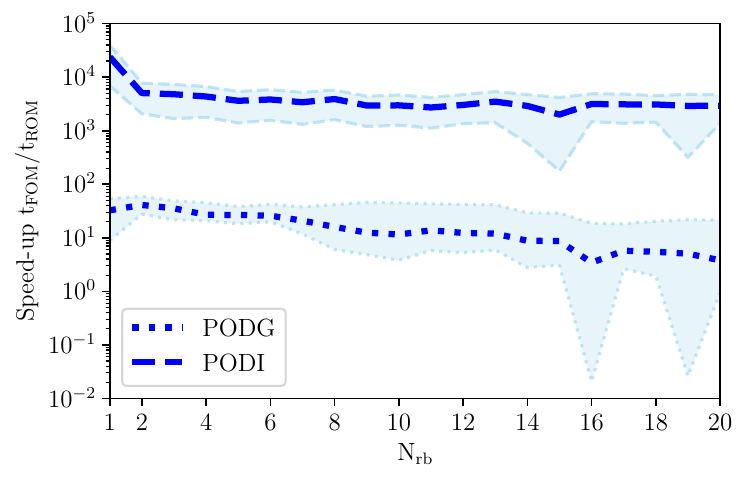}
    \caption{ }
    \label{fig:speedup_henkel_ins_only}
    \end{subfigure}
    \caption{Upper row: relative $L^2$-error across the test parameter set for Geometry 1 (a) and Geometry 2 (b). Lower row: speed-up across the test parameter set for Geometry 1 (c) and Geometry 2 (d). The bold line represents the mean value, while the shaded area indicates the range between minimum and maximum $L^2$-error and speed-up, respectively.}
    \label{fig:err_and_speedup_ins_only}
\end{figure}

In the following, the results of the two approaches are compared using identical test data for both methods. As outlined in Sec.~\ref{sec:intro}, the comparison focuses on the accuracy versus speed-up relation, the amount of data required by both methods, and the ability to extrapolate predictions outside the parameter range considered during training. The results for approximation accuracy as well as speed-up are shown in Fig.~\ref{fig:err_and_speedup_ins_only}. It is observed that the mean relative $L^2$-error rapidly decreases for both methods when increasing the size of the reduced basis. However, beyond a certain number of reduced basis functions, a plateau is reached at a very low error level, where adding more basis functions no longer improves the approximation quality. For the \gls{PODG} \gls{ROM}, this plateau aligns closely with the decay of the eigenvalues shown in Fig.~\ref{fig:eigvals_ins_only}. For \gls{PODI} however, the impact of interpolation is seen by the plateau being reached earlier. Due to its physics-based nature, \gls{PODG} leads to a significantly lower error compared to \gls{PODI}. On the other hand, when examining speed-up, it can be seen that \gls{PODI} consistently provides a high speed-up even with larger reduced basis sizes.

\begin{figure}
    \centering
    \begin{subfigure}[b]{0.47\textwidth}
    \hspace*{-1.25em}
    \includegraphics[width=\textwidth, keepaspectratio]{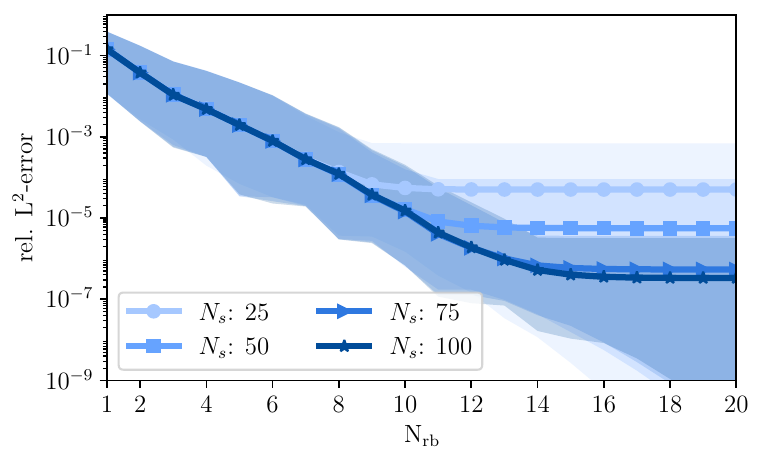}
    \caption{}
    \end{subfigure}
    \begin{subfigure}[b]{0.47\textwidth}
    \hspace*{-1.25em}
    \includegraphics[width=\textwidth, keepaspectratio]{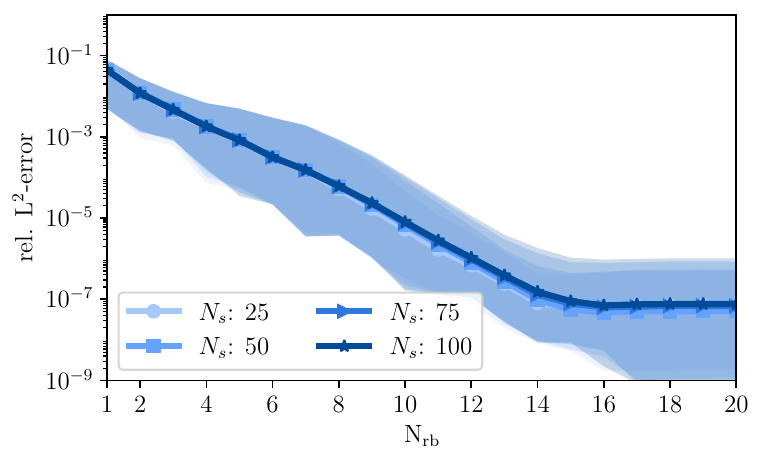}
    \caption{}
    \end{subfigure}
    \caption{Decay of the minimum, maximum and mean relative $L^2$-error across the test parameter set for different sizes of the snapshot set $\nsnap$ when setting up the \gls{ROM} by means of \gls{PODI} (a) and \gls{PODG} (b) for Geometry 2. The mean $L^2$-errors are indicated by the bold line, while the shaded areas represent the range between minimum and maximum $L^2$-error across the test parameter set.}
    \label{fig:data_hungriness}
\end{figure}

To further investigate how both methods behave when varying the amount of snapshot data that is provided during the offline phase, $\nsnap = \{25,50,75,100\}$ samples of the parameter space are considered. The resulting relative $L^2$-error is shown in Fig.~\ref{fig:data_hungriness}. It is evident that the \gls{ROM} constructed with the \gls{PODG} method approximates the \gls{FOM} very well even for very few snapshots. In contrast, the prediction accuracy of the \gls{ROM} constructed with \gls{PODI} shows to be highly sensitive to the amount of snapshot data when only a few data points are available, but becomes less sensitive as the training data increases. Hence, even with the maximum amount of data, \gls{PODI} still results in higher errors compared to \gls{PODG}. To evaluate the additional offline costs, well-performing configurations for both MOR methods are selected from the results in Fig.~\ref{fig:data_hungriness}. The analysis focuses on the costs of method-specific steps, excluding shared components such as the POD basis construction. Referring to the errors shown in Fig.~\ref{fig:data_hungriness}, a PODI ROM is chosen with $\nsnap = 75$ and $\romdim = 16$ and a PODG ROM with $\nsnap = 25$ and $\romdim = 16$. The tripled amount of snapshots leads to a respective increase in the snapshot generation time with approximately 220~\si{\s} for the 25 snapshots of~\gls{PODG} and approximately 734~\si{\s} for the 75 snapshots of~\gls{PODI}. Additional costs for~\gls{PODG} are offline costs for~\gls{DEIM} and for setting up the reduced operators, which add approximately 77~\si{\s} to the~\gls{PODG} offline costs. For~\gls{PODI}, the only additional step is the setup of the~~\gls{RBF} interpolation, which is completed in much less than 1~\si{\s} for the 16~\gls{POD} coefficients and 75 snapshots. The mean online run times can be quantified by 1.2~\si{\s} for~\gls{PODG} and very low 0.004~\si{\s} for~\gls{PODI}, which is reflected by the speed-up in Fig.~\ref{fig:speedup_henkel_ins_only}. Based on this analysis, a threshold of online evaluations is identified: For fewer than 366 evaluations,~\gls{PODG} is more efficient; for more than 366 evaluations,~\gls{PODI} becomes the more cost-effective choice in this example. Although the general trends are well captured by this evaluation, it should be noted that, excluding snapshot generation, two different tools were used. We remind the reader that their implementation can significantly affect the shown run times. All reported run times were computed on a system with an Intel i9-10980XE processor and 128 GB of RAM.

\begin{figure}
    \centering
    \includegraphics[scale=0.6]{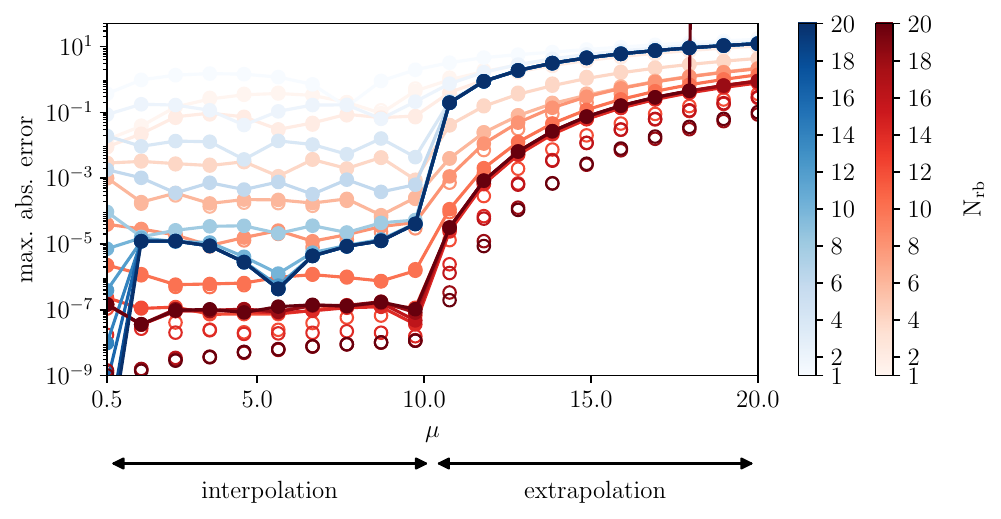}

    \caption{Comparison of interpolation and extrapolation capabilities of \gls{ROM}s for Geometry 2, based on 50 snapshots for $\wint \in \left[0.5~\si{\meter\per\second},10.15~\si{\meter\per\second}\right]$ and evaluated on 20 testing samples for $\wint \in \left[0.5~\si{\meter\per\second},20~\si{\meter\per\second}\right]$. For good interpretability, the maximum absolute error is shown over the parameter range for different sizes of the \gls{RB}. \gls{ROM} results are shown which are generated by \gls{PODG} with DEIM (red, filled marker), \gls{PODG} without \gls{DEIM} (red, empty marker) and \gls{PODI} (blue marker). The vertical line at $\wint \approx 18~\si{\meter\per\second}$ marks a test sample for which no solution was found with \gls{PODG} relying on \gls{DEIM}.}
    \label{fig:extrpolation_plot}
\end{figure}

Ultimately, the ability of the \gls{ROM}s to extrapolate to parameter values outside the trained range is investigated by splitting the training range so that both \gls{ROM}s are set up for inflow wind speeds $\wint \in [0.5~\si{\meter\per\second},10.15~\si{\meter\per\second}]$. However, they are still evaluated over the extended parameter range as before, i.e., $\wint \in [0.5~\si{\meter\per\second},20~\si{\meter\per\second}]$. For a better interpretability of the error, Fig.~\ref{fig:extrpolation_plot} shows the maximum absolute error over both the parameter range and the computational domain. It can be seen that both \gls{ROM}s provide small errors for parameters lying inside the parameter range for training. However, outside this range, the errors increase rapidly as the distance between the evaluated parameter and the trained range grows. Looking at the maximum absolute error at around $\wint = 11$ \si{\meter\per\second}, a jump is observed for both methods -- for \gls{PODG} the error increases by about two orders of magnitude, while for \gls{PODI}, the error jumps by about four orders of magnitude. Despite this, the prediction inaccuracies of both methods become more similar as the evaluated parameter moves further away from the training range. Comparing also the different RB sizes in Fig.~\ref{fig:extrpolation_plot}, it can be seen that a smaller RB can achieve similarly high extrapolation errors with \gls{PODI}. However, \gls{PODG} still benefits from a larger RB as seen before in Fig.~\ref{fig:mean_rel_l2_henkel_ins_only}, also within the extrapolation regime. However, even for a large RB both methods remain linear approaches approximating the nonlinear solution manifold of the \gls{FOM} based on data stemming from a lower velocity regime. Thus, if velocities increase above those considered within the training snapshots and the nonlinearity gets stronger, this approximation is increasingly likely to provide poorer results. Additionally, \gls{PODG} includes a \gls{DEIM} approximation for the nonlinear term. This approximation is, as shown in Sec.~\ref{sec:DEIM}, also set up based on training data. Thus, DEIM might fail as well to provide a good approximation if too far away from the trained parameter range. As also referred to in literature, for instance in~\cite{Cicci.2022}, insufficient \gls{DEIM} approximations can affect the convergence of the Newton algorithm solving the ROM in Eq.~\eqref{eq:POD_Galerkin_ROM}. This can be observed in Fig.~\ref{fig:extrpolation_plot} for the two highest velocity values and $\romdim = 20$. There, no solution could be found to the \gls{PODG} ROM including a \gls{DEIM} approximation for the nonlinear term. Since experiments showed that this is even the case with a doubled \gls{DEIM} basis size, Fig.~\ref{fig:extrpolation_plot} also includes results obtained by \gls{PODG} with an exact evaluation of the nonlinear term, i.e., without \gls{DEIM}. This is seen to (i) achieve overall better accuracy without the additional approximation and (ii) to perform more robustly for higher velocities within the extrapolation regime. However, removing the \gls{DEIM} approximation drastically increases the run time, making it infeasible for usage as a \gls{ROM}.

\subsubsection*{Discussion}
It has been shown that, although \gls{PODI} offers a greater flexibility and speed-up, the accuracy achieved by this purely data-driven method does not match the accuracy of the physics-based \gls{PODG}, even when using four times as much training data. Additionally, \gls{PODI} proves to be highly sensitive to insufficient training data, which causes a rapid decline in prediction accuracy. Regarding the ability to extrapolate to parameter values not included in the training snapshots, both methods perform poorly. Although it can be noted that \gls{PODG} performs slightly better when extrapolating close to the trained parameter range, a solution of the reduced system cannot be guaranteed further away from it if \gls{DEIM} is applied. 
Finally, two trade-offs can be identified when choosing between the two methods: First, the higher accuracy achieved by \gls{PODG} comes at the cost of a lower speed-up. This is observed from a higher, yet acceptable, error at a two to three orders of magnitude higher speed-up for the \gls{PODI} \gls{ROM}. Second, while the non-intrusiveness of \gls{PODI} offers great flexibility and straightforward implementation, it requires more data and yet provides results with lower accuracy, as shown in Fig.~\ref{fig:data_hungriness}. Nonetheless, the maximum relative $L^2$-error of \gls{PODI} still falls much below 1\% for $\nsnap = 75$ and $\romdim = 16$. Since, additionally, a high speed-up is of great interest for the application, the \gls{PODI} \gls{ROM} is extended to handle a system with two parameters -- both inflow wind velocity and wind direction.

\subsection{Two-parameter case}\label{sec:two_param_INS_only_ROM}
\begin{figure}
    \centering
    \begin{subfigure}[c]{0.47\textwidth}
    \hspace*{-1.25em}
    \includegraphics[width=\textwidth, keepaspectratio]{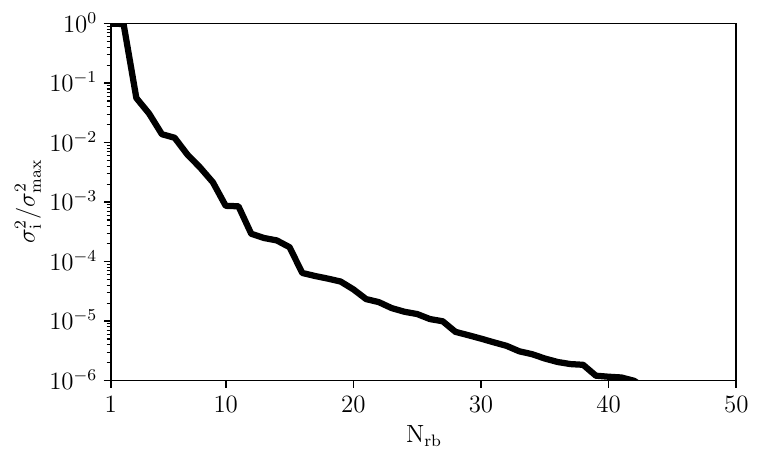}
    \caption{ }

    \end{subfigure}
    \begin{subfigure}[c]{0.47\textwidth}
    \hspace*{-1.25em}
    \includegraphics[width=\textwidth, keepaspectratio]{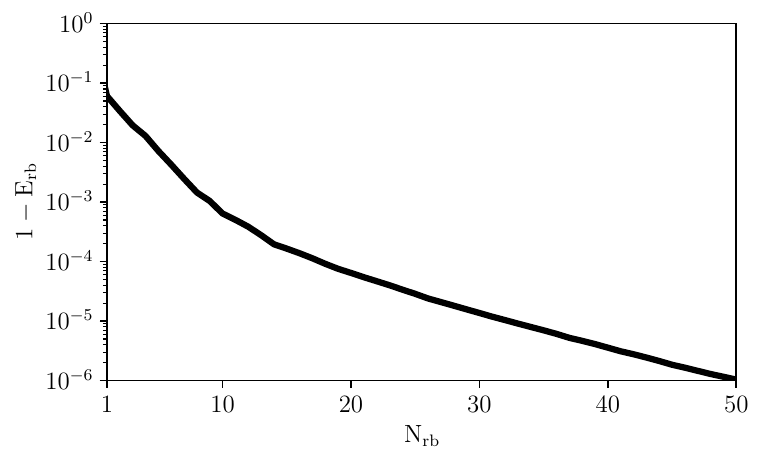}
    \caption{}

    \end{subfigure}
    \caption{Normalized eigenvalues of the correlation matrix (a) and retained energy (b) by a \gls{RB} of size $\romdim$ for a \gls{ROM} of the \gls{INS} problem parametrized by both inflow wind velocity and direction.}
    \label{fig:redbarkeit_ins_only_2params}
\end{figure}

\begin{figure}
    \centering
    \begin{subfigure}[b]{0.47\textwidth}
    \hspace*{-1.25em}
    \includegraphics[width=\textwidth, keepaspectratio]{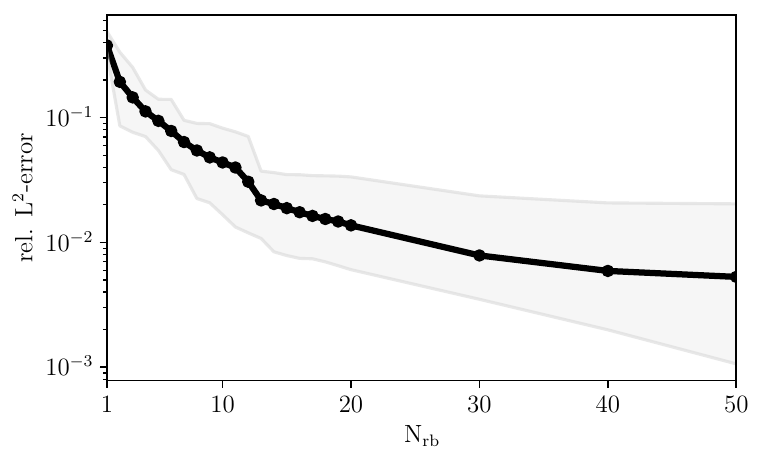}
    \caption{}
    \label{fig:mean_rel_l2_2param_ins_only}
    \end{subfigure}
    \begin{subfigure}[b]{0.47\textwidth}
    \hspace*{-1.25em}
    \includegraphics[width=\textwidth, keepaspectratio]{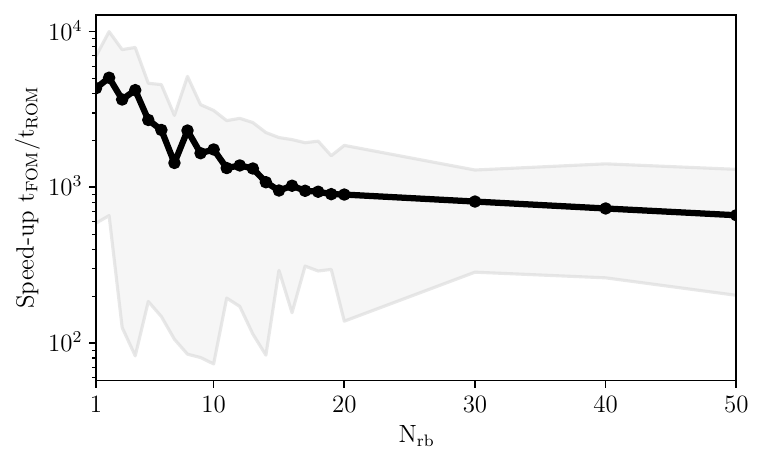}
    \caption{}
    \label{fig:speedup_over_n_2params_ins_only}
    \end{subfigure}
    \caption{Minimum, maximum and mean $L^2$-error (a) and speed-up (b) obtained by the \gls{PODI}-based \gls{ROM} developed to account for different inflow wind velocity and direction. The mean error and speed-up are indicated by the bold line, while shaded areas represent the range between minimum and maximum error or speed-up across the test parameter set.}
    \label{fig:err_and_speedup_2params_ins_only}
\end{figure}

Building on the results from the previous study, the \gls{PODI}-based \gls{ROM} is extended to include a second parameter -- the wind direction. Here, a refined version of the mesh shown in Fig.~\ref{fig:mesh_geom2}, with approximately $1.7\times 10^5$ degrees of freedom, is used as computational domain. The Dirichlet boundary condition is now defined as $\mbg_D = \wint[\cos(\wdir),\sin(\wdir)]^\intercal$ with the inflow wind velocity $\wint$ in \si{\meter\per\second} and the direction $\wdir$ in radians \cite{Gioia.2024}. Consequently, the \gls{INS} problem in Eq.~\eqref{eq:INS_strong} is transformed into an enclosed flow problem, as treated in \cite[Sec.~9.3.5]{Elman.2014}, without a Neumann boundary condition. As before, the reducibility of the problem is assessed by evaluating the eigenvalues of the correlation matrix and the retained energy for the \gls{RB} size $\romdim$, as shown in Fig.~\ref{fig:redbarkeit_ins_only_2params}. It can be observed that the decay of the eigenvalues is slower than in the previous case, where only wind intensity has been considered as a parameter.
This suggests an increased need for a larger \gls{RB}. The parameters of the \gls{ROM} are summarized in Table~\ref{tab:Parameters_MOR_2params_ins}. For the Reynolds number of the two-parameter case, the diameter of the refined mesh around the region of interest, as it can be seen in Fig.~\ref{fig:mesh_geom2}, is chosen as the characteristic length for the Reynolds number.
The resulting approximation accuracy and the speed-up are presented in Fig.~\ref{fig:err_and_speedup_2params_ins_only}. As indicated by the eigenvalues of the correlation matrix, introducing the second parameter makes the problem significantly harder to reduce. Hence, the mean $L^2$-error is not as low as in the previous case but remains below $2\%$ for a \gls{RB} size of twenty. Additionally, a high speed-up of approximately 1000 is achieved by the \gls{ROM}.

\begin{table}
\caption{Parameters of the full-order and reduced-order model featuring a parametric dependence on the inflow wind velocity and direction.}
\begin{center}
\begin{tabular}{c c c}
\toprule
\multicolumn{2}{c}{\textbf{Parameter}} & \textbf{Value} \\ \midrule
\multicolumn{2}{c}{Degrees of freedom} & 169,462 \\
\hline
\multicolumn{2}{c}{Reynolds number} & $\Rey \in \left[0,220\right]$ \\
\hline
\multicolumn{2}{c}{Inflow wind velocity $(\si{\meter\per\second})$} &  $\wint \in \left[1e-4,12\right]$ \\
\hline
\multicolumn{2}{c}{Inflow wind direction $(\si{\deg})$} & $\wdir \in \left[0,360\right]$ \\ \hline
\multicolumn{2}{c}{Subspace \gls{POD} ($\romdim$)}& $\romdim \in [1,50]$\\ \hline
\multirow{2}*{Snapshots} & wind intensity $\wint$ & 20 \\ \cline{2-3}
  & wind direction $\wdir$ & 150 \\ \hline
\multirow{2}*{Test set size} & wind intensity $\wint$ & 4 \\ \cline{2-3}
 & wind direction $\wdir$ & 30 \\ \bottomrule
\end{tabular}
\label{tab:Parameters_MOR_2params_ins}
\end{center}
\end{table}

\section{Application Case for Crisis Management - Gas Leakage Incident at a Chemical Plant}\label{sec:application}

\subsection{Uncertainty quantification for wind measurement errors}\label{sec:uq}
The following scenario is based on the application presented in \cite{Schneider.2025}, where an Emergency Response Inference Mapping (ERIMap) is introduced. The goal is to enable a systematic assessment of emergency situations tailored to support decision-making. To illustrate the approach, a chlorine gas leak at a chemical plant located in D\"usseldorf, Germany, is considered. As a model for the dispersion of chlorine gas a Gaussian plume model implemented in ALOHA \cite{jones.2013} is applied. In this work, the modeling approach is extended by describing the plume development by an advection-diffusion problem instead. To facilitate rapid evaluation during emergencies and enable multi-query analysis, the \gls{ROM} approach presented in Sec.~\ref{sec:two_param_INS_only_ROM} is employed to predict the wind field. This way, only the computationally less expensive advection-diffusion problem needs to be solved. To demonstrate the advantages of this approach, the influence of uncertainties in wind measurements is explored by using a Monte Carlo analysis~\cite{Fishman.1996} with 2,000 samples implemented in the open-source multi-query analysis framework \textit{QUEENS}~\cite{Biehler.2026}. For this analysis, the mean wind velocity and direction values are derived from one year of wind data collected for D\"usseldorf by the German weather service (DWD)~\cite{DWD.CDC}. The inflow wind velocity and direction are assumed to vary uniformly within a specific range around the annual mean. For the inflow wind speed, an uncertainty of $\pm 5\%$ is assumed, corresponding to an absolute range of $\wint = 4.1 \pm 0.2~\si{\meter\per\second}$. For the wind direction, a $\pm 5$~\si{\deg} range is employed which translates to $\wdir = 191 \pm 5~\si{\deg}$, where $0~\si{\deg}$ denotes wind coming from the North.

\begin{figure}
    \centering
    \begin{subfigure}[c]{0.47\textwidth}
    \hspace*{0.75em}
    \includegraphics[width=\textwidth, keepaspectratio]{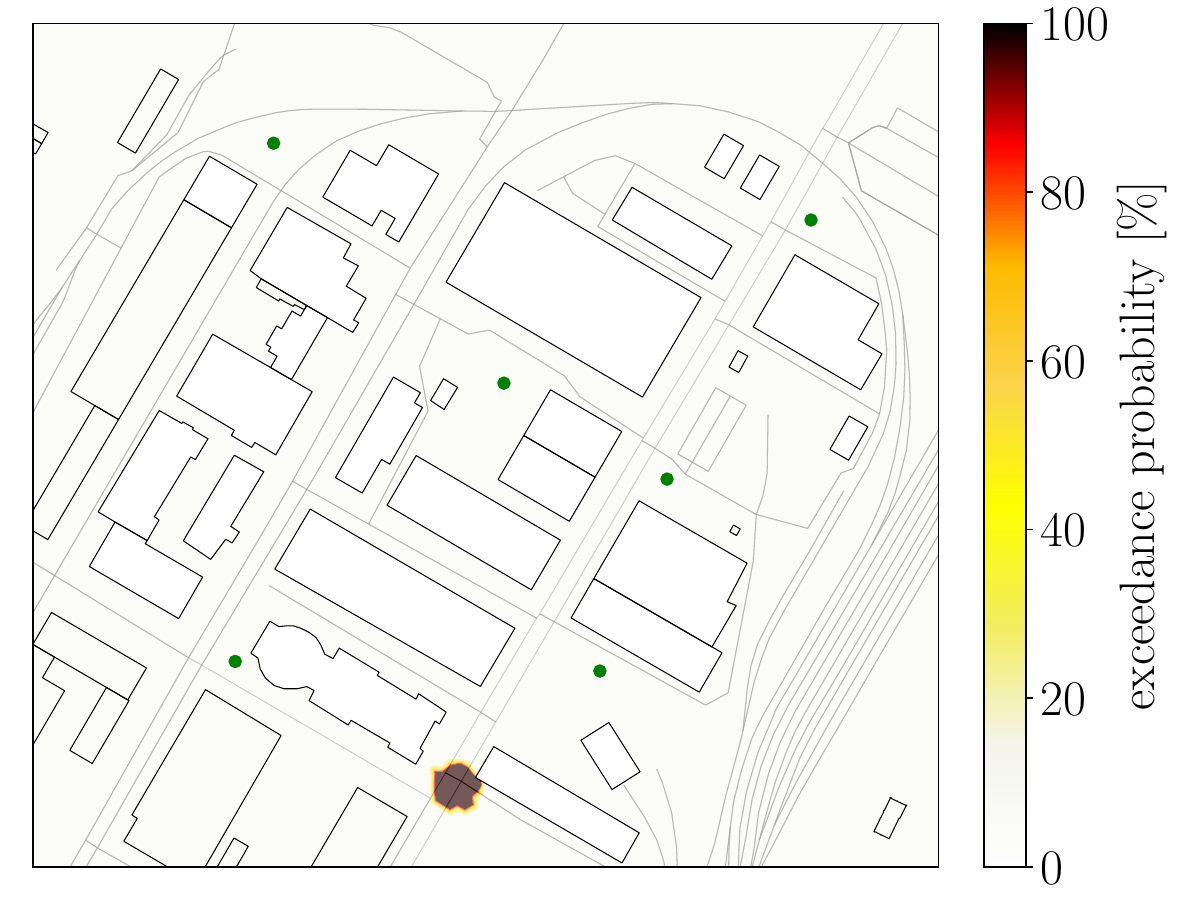}
    \caption{ }
    \end{subfigure}
    \begin{subfigure}[c]{0.47\textwidth}
    \hspace*{0.75em}
    \includegraphics[width=\textwidth, keepaspectratio]{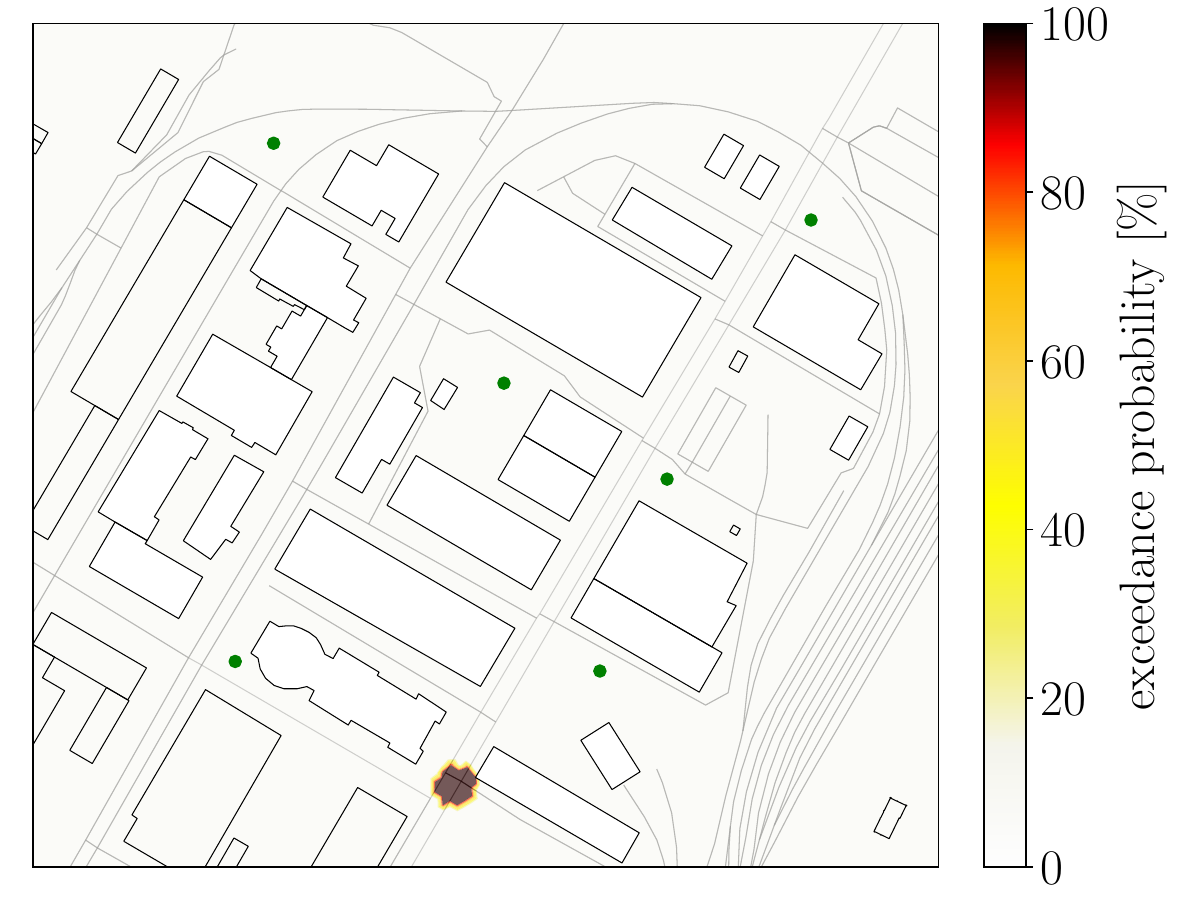}
    \caption{ }
    \end{subfigure}
    \begin{subfigure}[c]{0.47\textwidth}
    \hspace*{0.75em}
    \includegraphics[width=\textwidth, keepaspectratio]{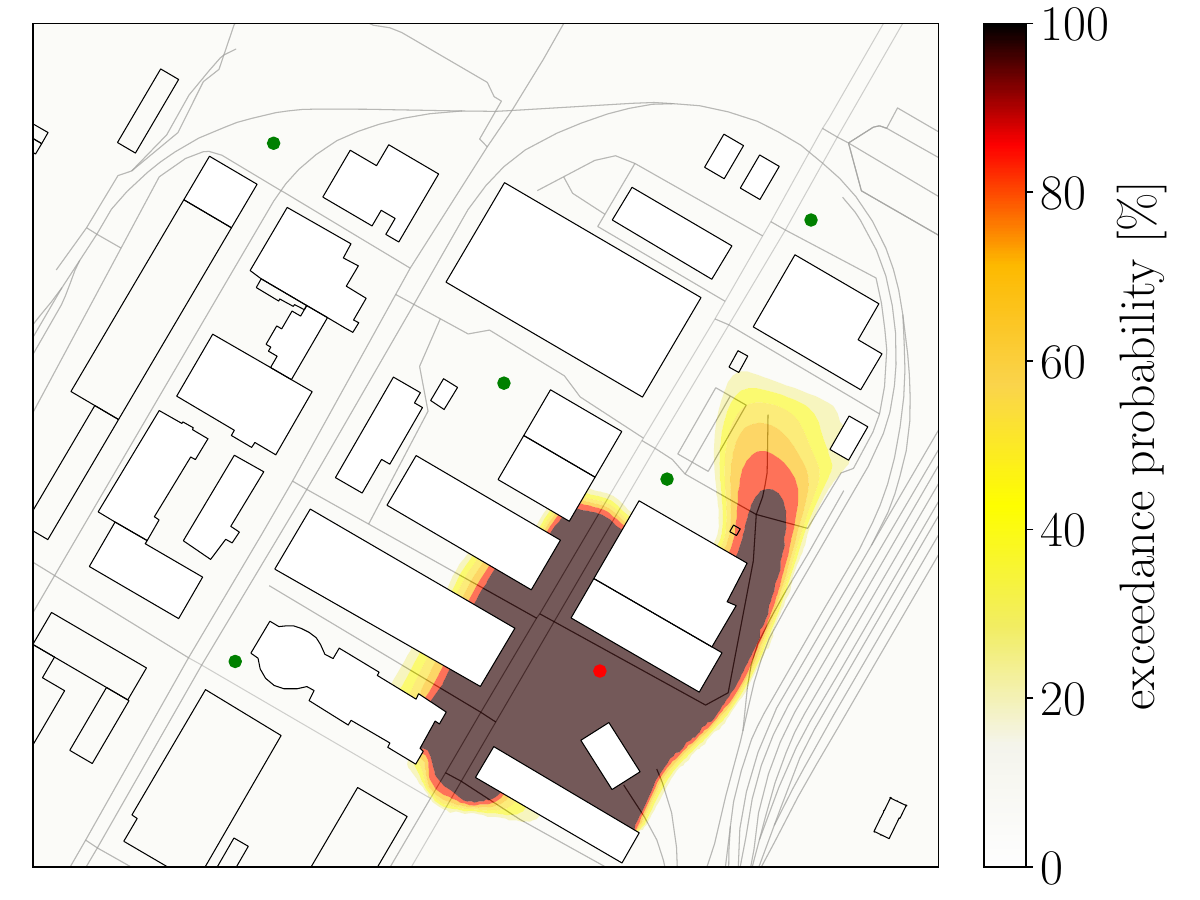}
    \caption{ }
    \end{subfigure}
    \begin{subfigure}[c]{0.47\textwidth}
    \hspace*{0.75em}
    \includegraphics[width=\textwidth, keepaspectratio]{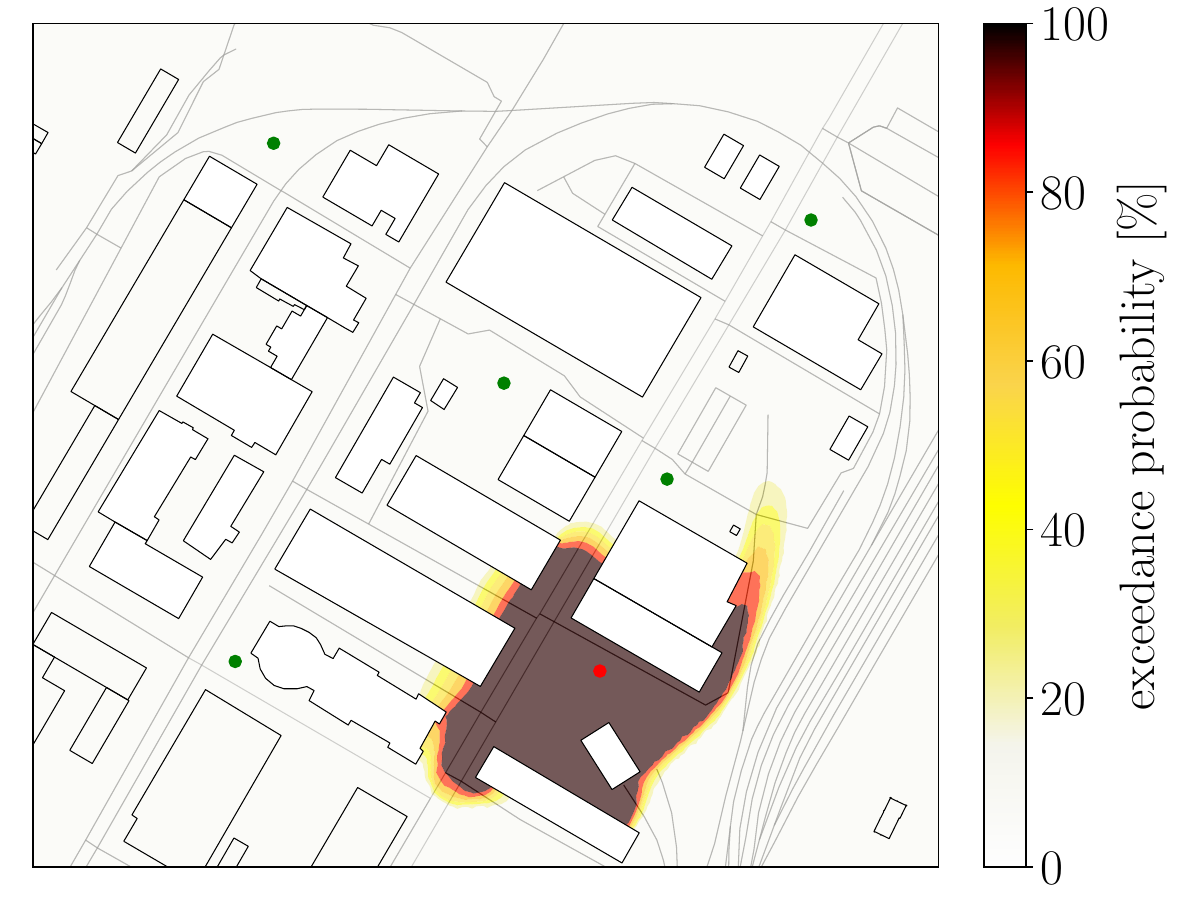}
    \caption{ }
    \end{subfigure}
    \begin{subfigure}[c]{0.47\textwidth}
    \hspace*{0.75em}
    \includegraphics[width=\textwidth, keepaspectratio]{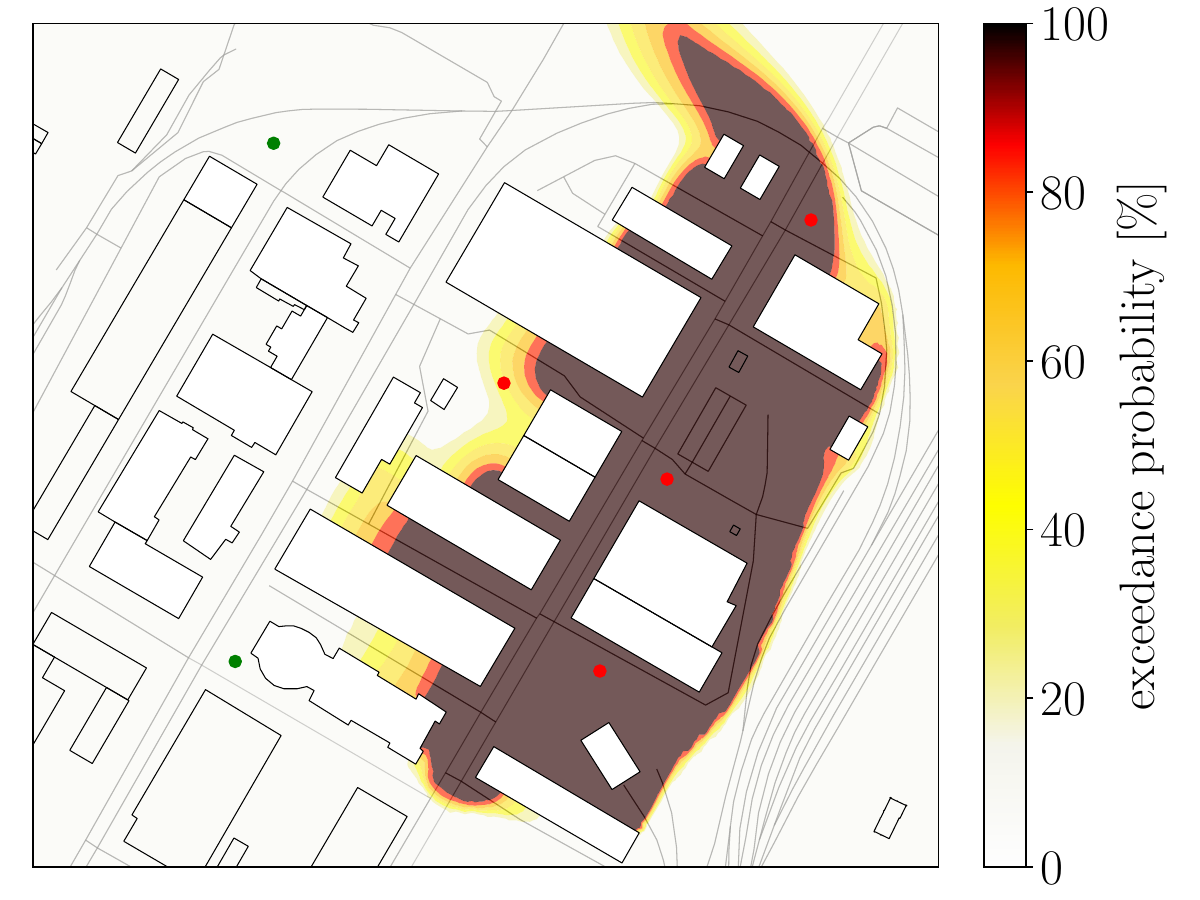}
    \caption{ }
    \end{subfigure}
    \begin{subfigure}[c]{0.47\textwidth}
    \hspace*{0.75em}
    \includegraphics[width=\textwidth, keepaspectratio]{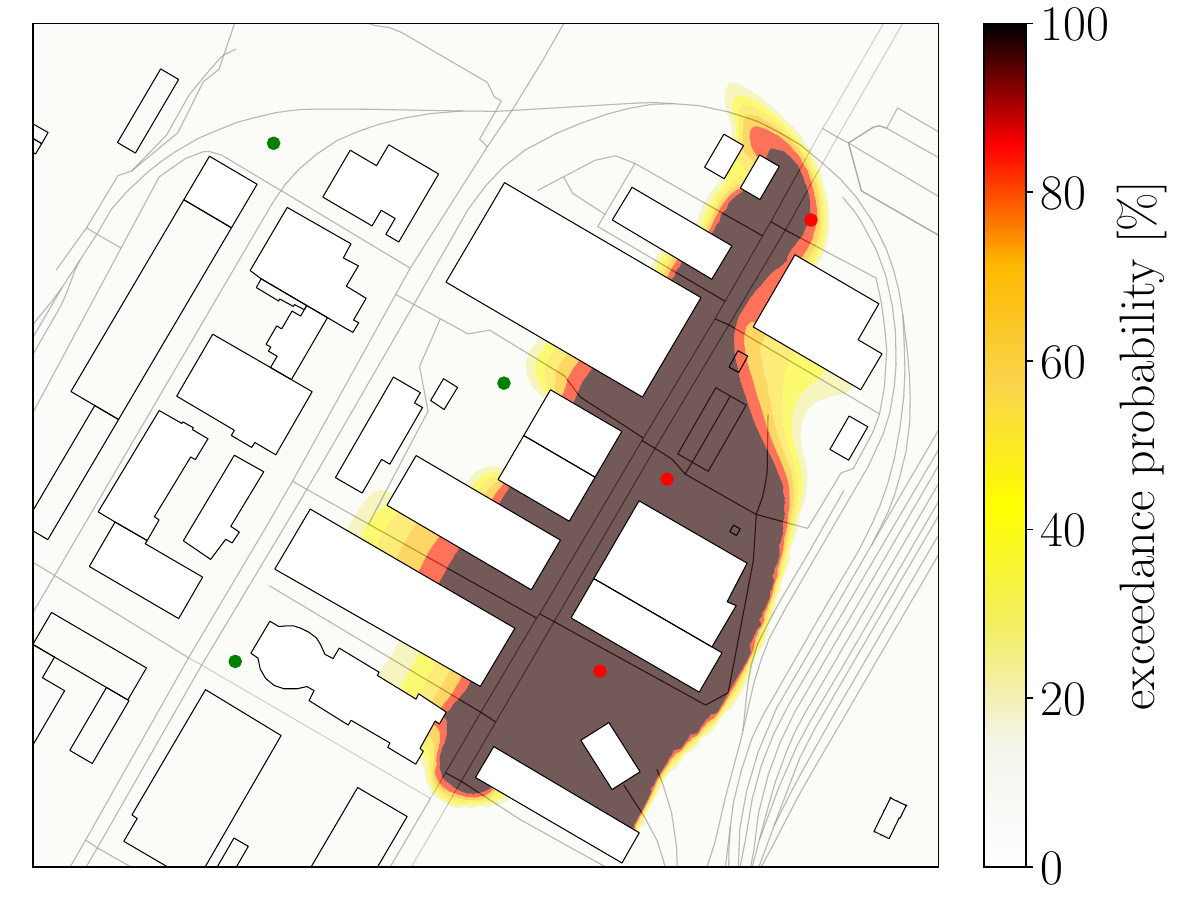}
    \caption{}
    \end{subfigure}
    \caption{Exceedance maps showing probabilities of exceeding concentration ratios above $0.05$ (left column) and $0.1$ (right column) based on the Monte Carlo analysis for simulation times $t  = 0$~\si{\s} (upper row), $t=10$~\si{\min} (middle row), $t=30$~\si{\min} (lower row). Green and red dots indicate (un-)available assembly points.}
    \label{fig:ex_prob}
\end{figure}

\begin{figure}
    \centering
    \begin{subfigure}[b]{0.47\textwidth}
    \hspace*{-1.25em}
    \includegraphics[width=\textwidth,keepaspectratio]{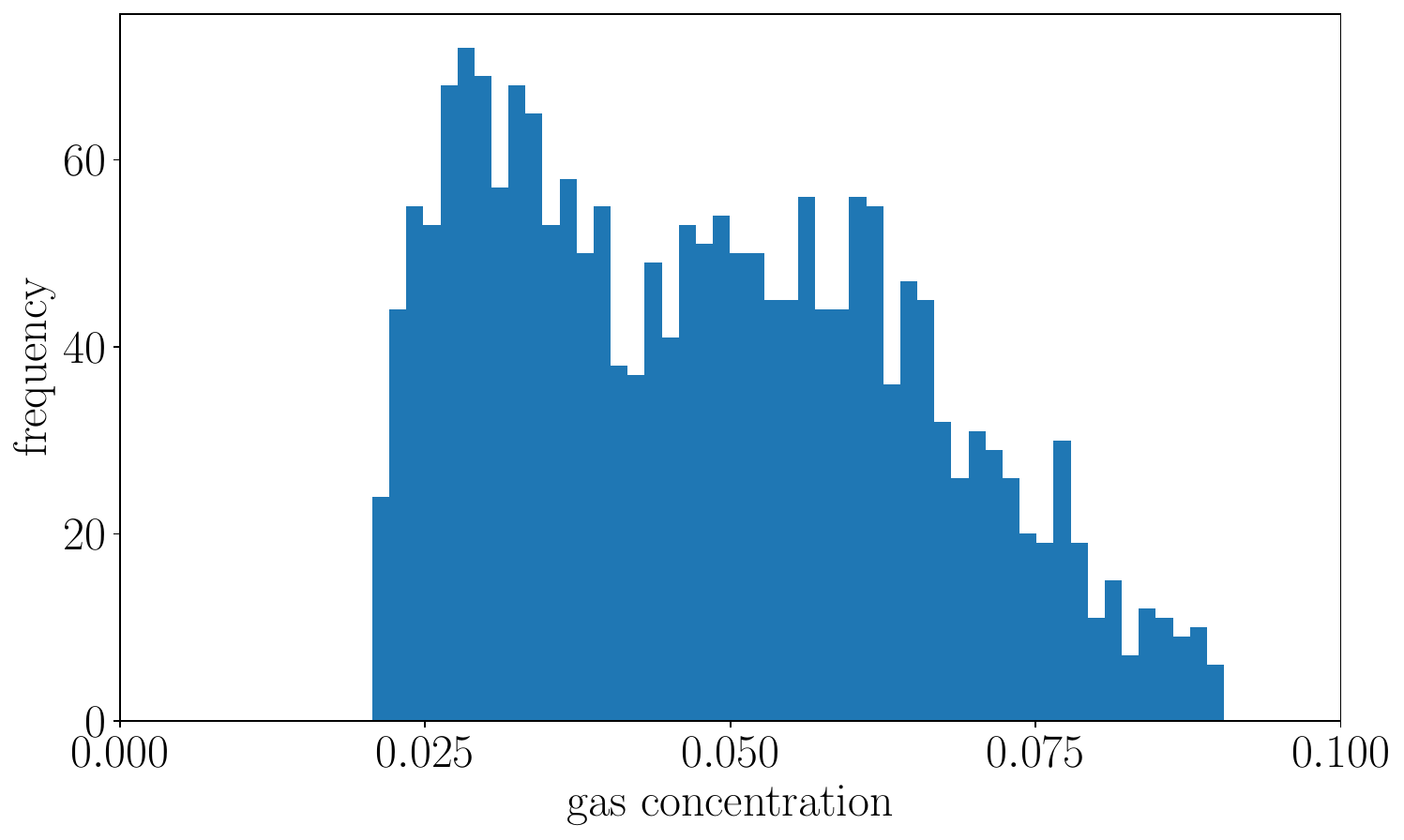}
    \caption{ }
    \end{subfigure}
    \begin{subfigure}[b]{0.47\textwidth}
    \hspace*{-1.25em}
    \includegraphics[width=\textwidth, keepaspectratio]{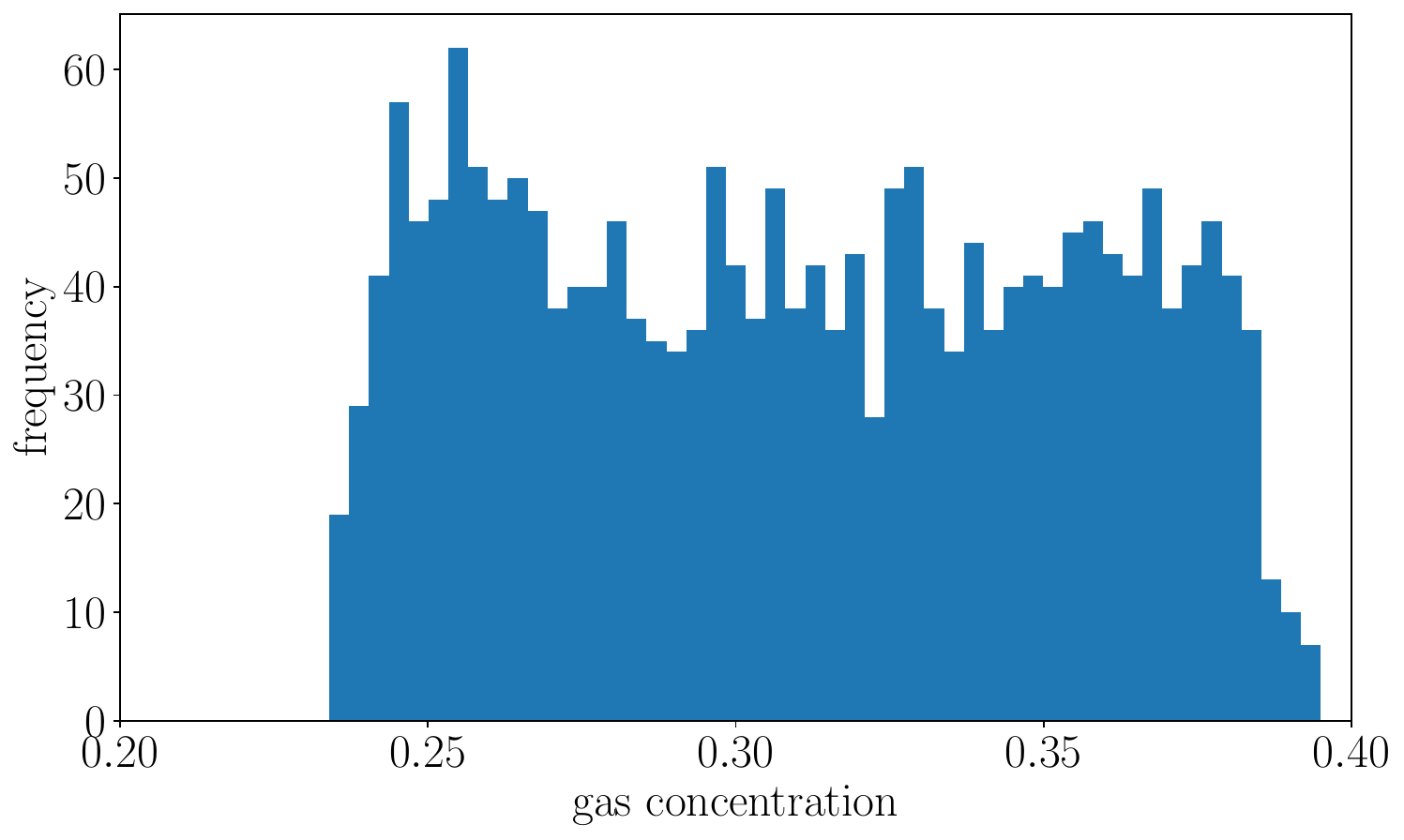}
    \caption{ }
    \end{subfigure}
    
    \caption{Distribution of gas concentration values at two selected assembly points in the middle of the domain (a) and the lower right (b) at $t = 30$~\si{\min} stemming from uncertainty in the wind velocity and direction measurements.}
    \label{fig:nodal_values_MC_most_var}
\end{figure}

To ensure reasonable computational times while maintaining accurate wind field predictions, the \gls{ROM} for the \gls{INS} problem, as presented in Sec.~\ref{sec:two_param_INS_only_ROM}, is applied with $\romdim = 20$. For the UQ range, a very good overall approximation accuracy could be observed for the \gls{ROM} with a maximum relative $L^2$-error of $8.9\times 10^{-3}$, a mean relative $L^2$-error of $8.7\times 10^{-3}$, and a minimum relative $L^2$-error of $8.4\times 10^{-3}$. Using the resulting wind field predictions, the advection-diffusion problem is solved for $t \in \left[0~\si{\min},30~\si{\min}\right]$. The resulting probability to exceed a certain threshold for the concentration value -- here $0.05$ and $0.1$ -- are shown in Fig.~\ref{fig:ex_prob} for the time steps $t\in \{0~\si{\min},10\si{\min},30\si{\min}\}$. High probabilities indicate that the concentration value at the respective location exceeds the threshold concentration for many Monte Carlo simulations. In addition to the buildings of the industrial site, the paths and railway lines are shown, together with assumed assembly points, in order to demonstrate feasibility of the plots for assessing the availability and safety of escape routes and meeting points. For further insights,  Fig.~\ref{fig:nodal_values_MC_most_var} shows the distribution of the concentration values at $t = 30~\si{\min}$ for two of the assumed assembly points, one with an exceedance probability of $100$\% and one with a exceedance probability around $50$\% for a threshold value of $0.05$. At both assembly points, the concentration values vary widely, each following a different distribution. Thus, a description in terms of mean and standard deviation alone would provide an incomplete description of the nodal concentration probability density distributions.
Although the \gls{ROM} accelerates such analyses and shows to be very accurate in our case, it has to be kept in mind that it introduces approximation errors to the system. An estimation of the error made by the applied \gls{ROM} for the different parameter values within the uncertainty range can be obtained by evaluating the \gls{ROM} performance across the testing parameter space as shown in Fig.~\ref{fig:mean_rel_l2_2param_ins_only} as well as it was performed for the error measures reported above. This error estimation can be further improved by the spatial distribution of the prediction error, providing additional information on how these errors might affect the region of interest.

\begin{figure}
    \centering
    \begin{subfigure}[b]{\textwidth}
    \centering
    \includegraphics[width=0.65\textwidth,keepaspectratio]{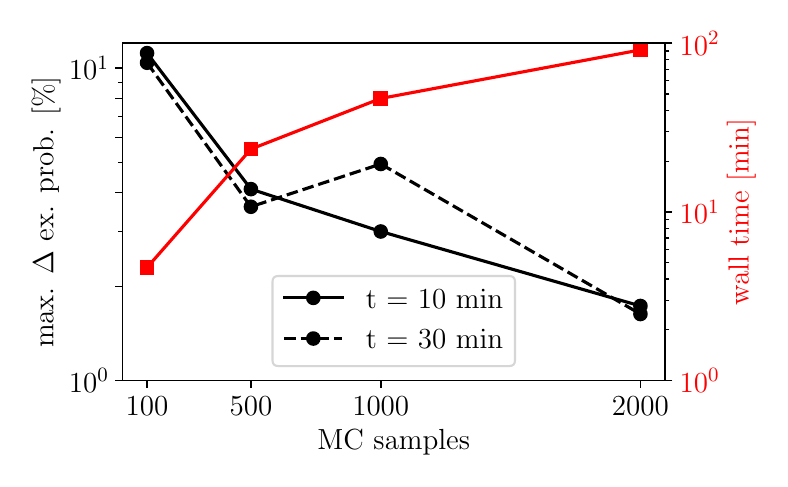}
    \caption{ }
    \end{subfigure}
    \begin{subfigure}[b]{0.47\textwidth}
    \hspace*{0.75em}
    \includegraphics[width=\textwidth, keepaspectratio]{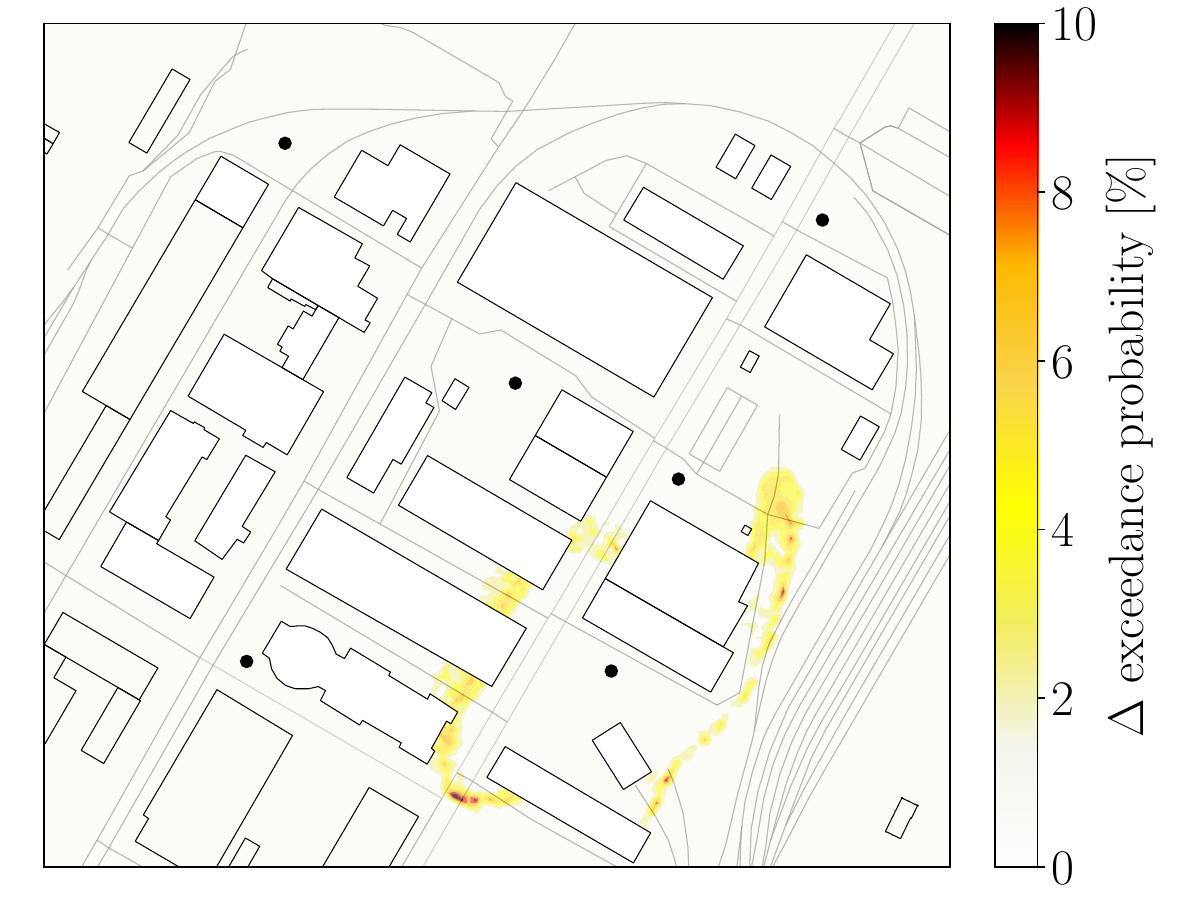}
    \caption{ }
    \end{subfigure}
    \begin{subfigure}[b]{0.47\textwidth}
    \hspace*{0.75em}
    \includegraphics[width=\textwidth, keepaspectratio]{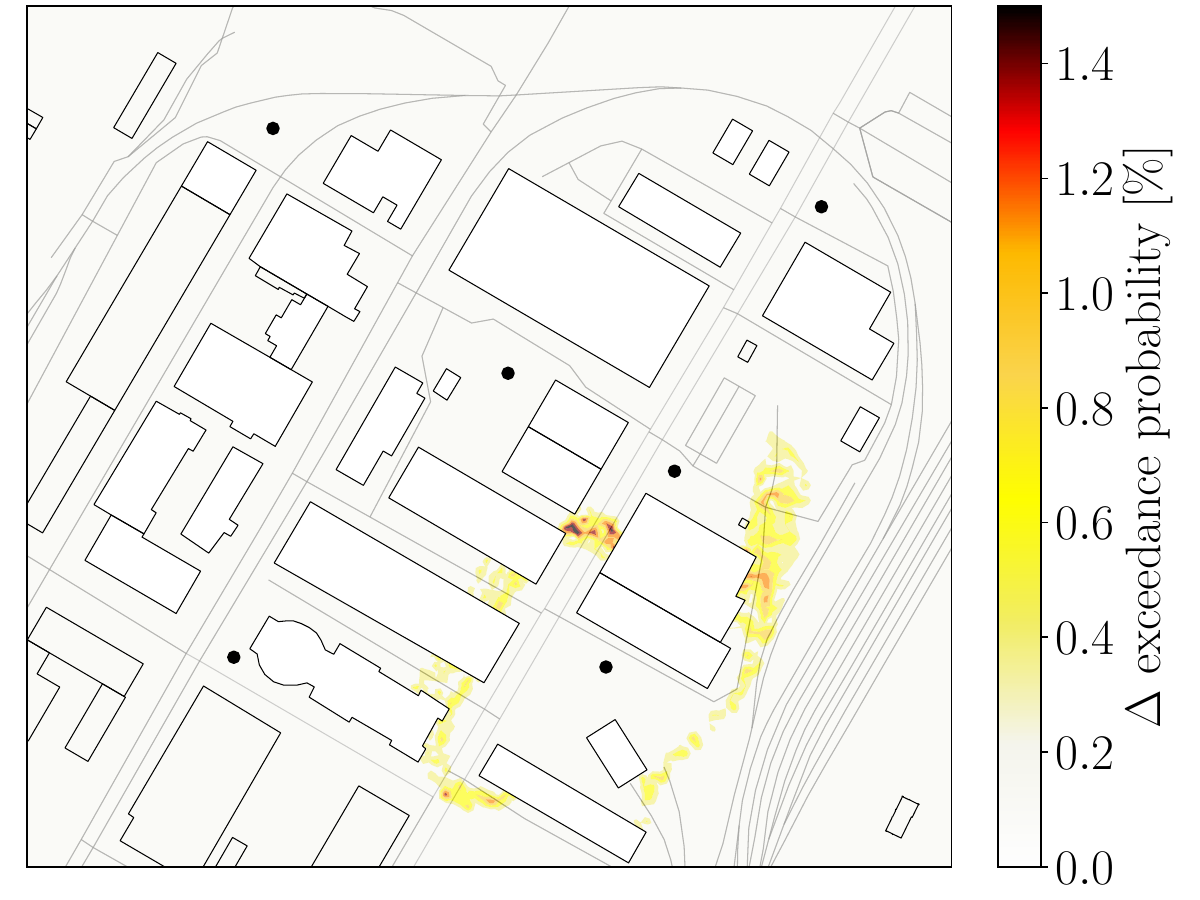}
    \caption{ }
    \end{subfigure}
    \begin{subfigure}[b]{0.47\textwidth}
    \hspace*{0.75em}
    \includegraphics[width=\textwidth, keepaspectratio]{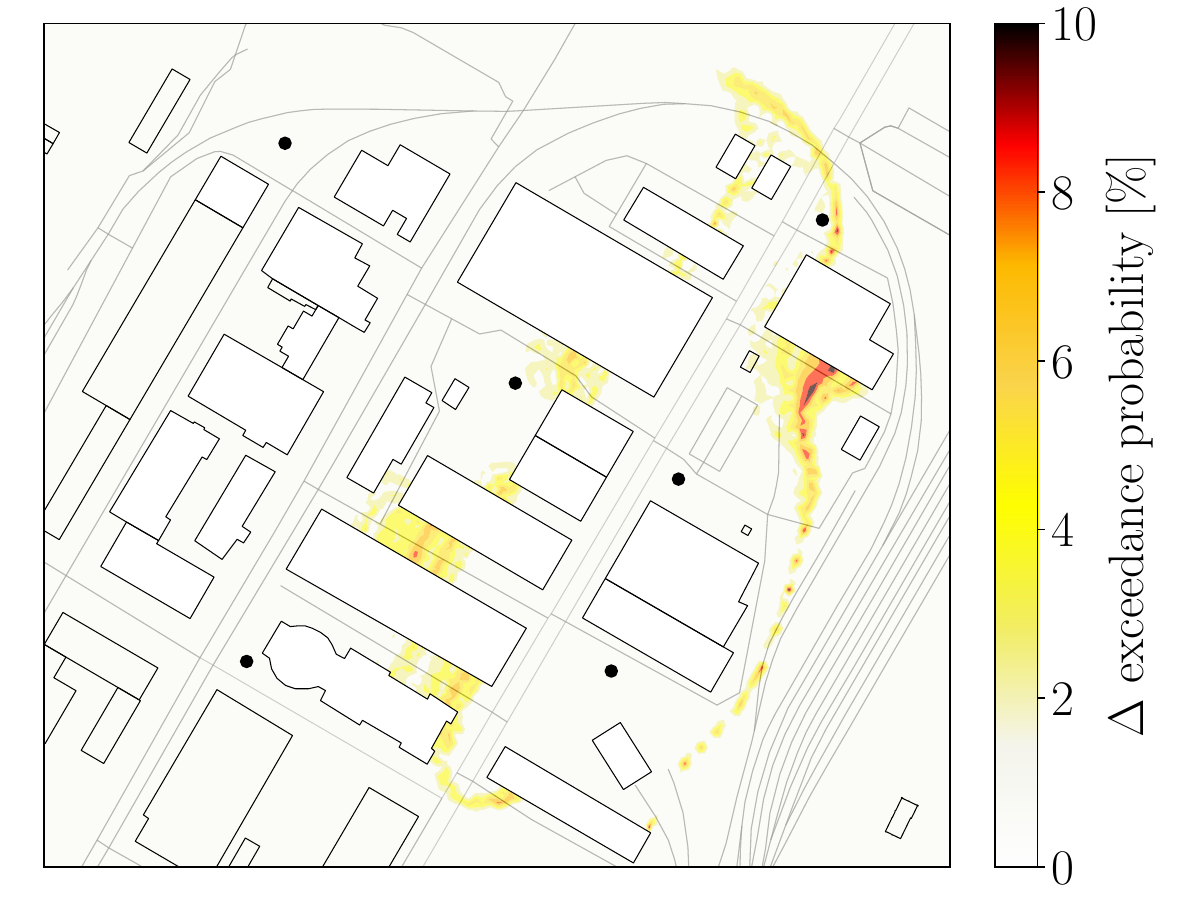}
    \caption{ }
    \end{subfigure}
    \begin{subfigure}[b]{0.47\textwidth}
    \hspace*{0.75em}
    \includegraphics[width=\textwidth, keepaspectratio]{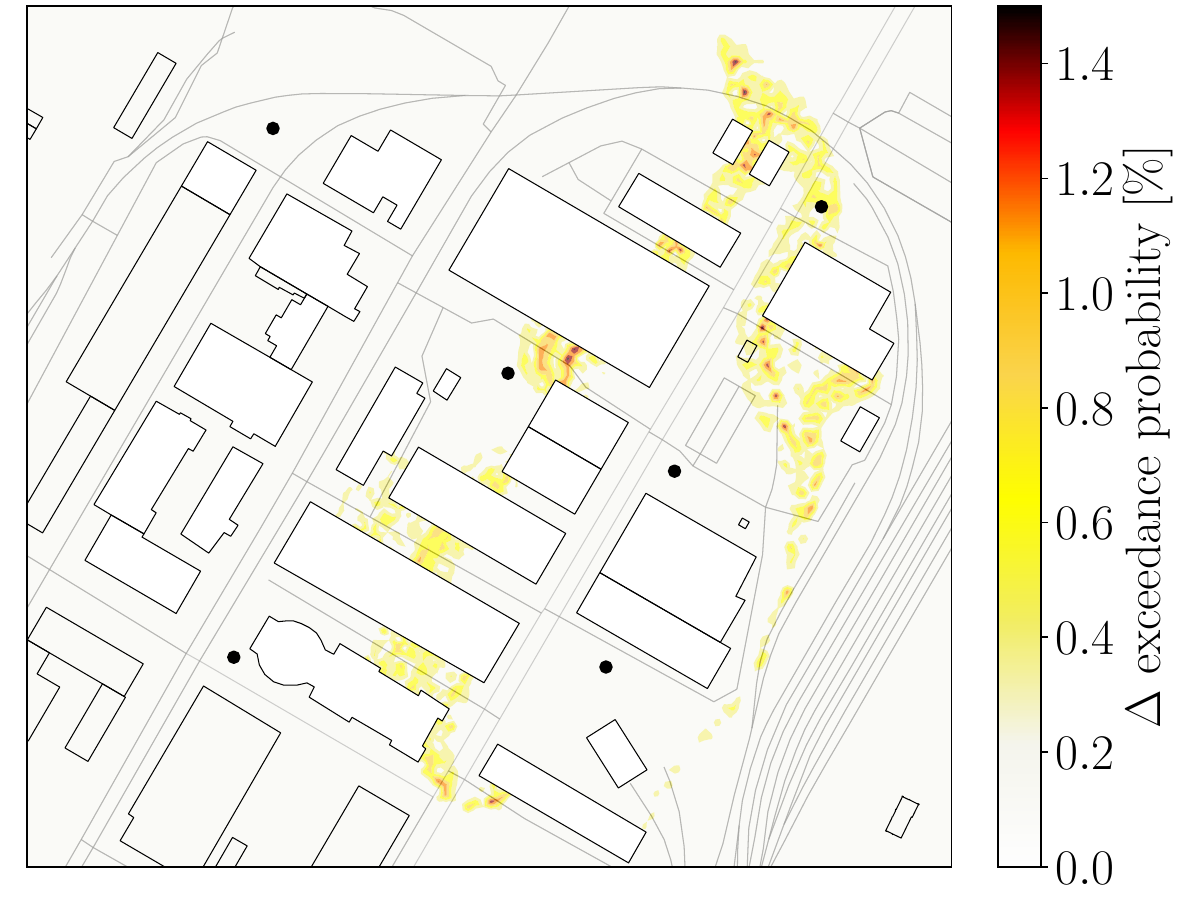}
    \caption{ }
    \end{subfigure}
    
    \caption{Maximum absolute difference in the exceedance probability computed against probabilities obtained for 3,000 Monte Carlo samples, jointly displayed with respective wall times for the Monte Carlo analysis in (a); spatial distribution for $t = 10$~\si{\min} (middle row), $t=30$~\si{\min} (lower row) of absolute differences in the exceedance probabilities of 100 samples compared to 3,000 samples (b) and (d), as well as 2,000 compared to 3,000 Monte Carlo samples (c) and (e).}
    \label{fig:mc_sampels}
\end{figure}

The here presented results for 2,000 Monte Carlo samples correspond to a total computation time of approximately 90~\si{\min} to compute the different solutions for the concentration field for $t \in [0~\si{\min},30~\si{\min}]$ . However, the run time heavily depends on the amount of considered Monte Carlo samples. If less samples are used, computation times can go down to, for instance, $4$~\si{\min} for 100 Monte Carlo samples. However, changes in the exceedance probability have to be carefully evaluated when varying the amount of samples. Therefore, Fig.~\ref{fig:mc_sampels} displays curves of maximum absolute difference in the exceedance probability with a threshold of $0.1$ as well as the spatial distribution of the absolute differences when comparing Monte Carlo results for 100, 2000 and 3000 samples.

\subsection{Visualization in interactive dashboard}\label{sec:dashboard}
\begin{figure}
    \centering
    \includegraphics[width=0.9\textwidth, keepaspectratio]{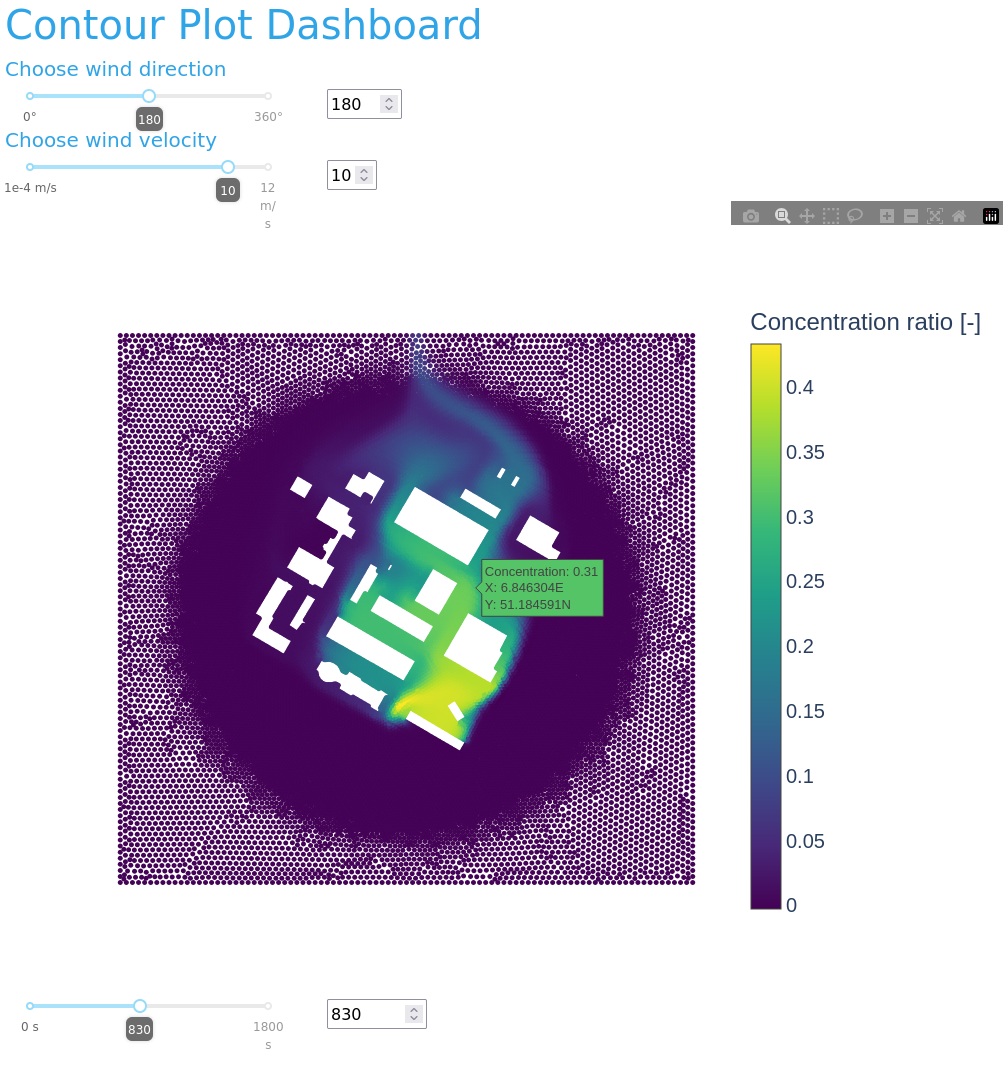}
    \caption{Screenshot of the interactive dashboard displaying fast online evaluations for user-specified parameter values.}
    \label{fig:dashboard}
\end{figure}

To enable the practical use of the computed data, the reduced model can be integrated into a digital twin framework and results can be visualized through an interactive dashboard, as shown in Fig.~\ref{fig:dashboard}. Here, users can easily adjust the inflow wind direction and velocity, as well as the time step, directly from the graphical user interface. The \gls{ROM} then allows for a rapid evaluation of the spatio-temporal concentration field based on the user input. The latency times for this computation and the subsequent visualization are summarized in Table~\ref{tab:dashboard}. The concentration field is displayed in such a way that concentration values and the respective local information can be easily accessed by hovering over the specific regions. This can be particularly helpful in emergency situations, for instance, when coordinating evacuation teams is necessary, and can also serve as a valuable tool for training exercises. Although the example shown here represents an initial step, the visualization demonstrates how the developed \gls{ROM} can be effectively used within the context of crisis management. Furthermore, it highlights the benefits of embedding simulations into such frameworks, offering an intuitive interface for real-time decision-making. 

\begin{table}
\caption{Computational times of dashboard operations for $t = 30$~\si{\min} at a time step of $\Delta t = 5$~\si{\s}. Shown are (i) the evaluation time of the INS ROM, (ii) the post-processing of the ROM results for subsequent evaluation of the AD problem, (iii) the mean evaluation time per time step of the AD problem, and (iv) the latency time for plotting the result on the dashboard. All timings are averaged over 100 runs for random choices of $\wint$ and $\wdir$.}
\begin{center}
\begin{tabular}{c @{\hspace{1cm}} c @{\hspace{1cm}} c @{\hspace{1cm}} c}
\toprule
\textbf{$\tau_{INS, ROM}$} & \textbf{$\tau_{ROM, post}$} & \textbf{$\tau_{AD, FOM}$} & \textbf{$\tau_{plot}$}\\ \midrule
$\sim$ 0.006~\si{\s} & $\sim$ 0.17~\si{\s} & $\sim$ 0.01~\si{\s} & $\sim$ 0.13~\si{\s} \\
\bottomrule
\end{tabular}
\label{tab:dashboard}
\end{center}
\end{table}
\section{Conclusion and Outlook}\label{sec:conclusion}
In this paper, a comprehensive comparison between \gls{PODI} and \gls{PODG} has been conducted for predicting a parametrized laminar wind field for different inflow wind velocities. The use case is motivated by the scenario of airborne contaminant transport across a built-up area, modeled as a two-dimensional domain with real-world building layouts.
The comparison has revealed that both methods provide sufficiently accurate results, though a clear trade-off between approximation accuracy and speed-up has been observed. Furthermore, another trade-off has been identified between the amount of data required to set up the \gls{ROM} and the accuracy loss that one is willing to accept for the benefits of non-intrusiveness. Given that \gls{PODI} has still led to a maximum relative $L^2$-error of much below 1\%, combined with its significant speed-up, the approach has been subsequently extended to account for two parameters -- the inflow wind direction and velocity. This extension allows for more realistic modeling. Additionally, potential benefits of a reduced model in the considered scenario have been demonstrated by performing a Monte Carlo analysis, which can help identifying areas with consistent, possible and expected gas concentration based on uncertainties in wind measurements. Ultimately, the computed information could be exploited in an easily accessible manner by embedding the \gls{ROM}-based numerical simulation into an interactive dashboard that visualizes contaminated areas for varying, user-specified wind conditions.\\
As mentioned, the presented numerical setup focuses on a two-dimensional representation of a real-world geometry, considering only a laminar, two-dimensional flow. Although this simplification enabled an efficient comparison by keeping costs low for \gls{FOM} evaluations, it can only serve as a starting point towards more realistic three-dimensional, turbulent flows. Therefore, future work will aim to extend this study, including the presented comparison, to three-dimensional flows and higher Reynolds numbers. A common approach for modeling urban air flows is to employ Reynolds-averaged Navier-Stokes (\gls{RANS}) equations or large eddy simulations (\gls{LES})~\cite{Blocken.2018,GarciaSanchez.2014,Owkes.2025,Ding.2025,Teng.2025}. Applications of \gls{PODI} for the two modeling approaches can be found, for instance, in~\cite{Khamlich.2023,Gadalla.2021,Masoumi.2022}. Moreover, since both methods, \gls{PODI} and \gls{PODG}, rely on a linear subspace, the evaluation of nonlinear \gls{MOR} approaches will be an additional aspect of future research. Lastly, setting up a \gls{ROM} for the wind field yielded certain advantages such as reduced constraints on source locations or emission rates compared to a ROM for the coupled problem. Such a \gls{ROM} would enable a reduced solution for the concentration field directly, but is specifically designed to consider selected parameters. Still, introducing a reduced model for the \gls{AD} problem or the coupled problem is another aspect of high interest to further accelerate the evaluation process.

\section*{Acknowledgment}
A.P. acknowledges support by \textit{dtec.bw - Digitalization and Technology Research Center of the Bundeswehr} (project \textit{RISK.twin}). dtec.bw is funded by the European Union - NextGenerationEU. 

\glsaddall
\printnoidxglossaries

\bibliographystyle{unsrtnat}

\bibliography{references}

\end{document}